\documentclass[onecolumn]{aa}

\usepackage{aabib}

\input{psfig.sty}

\begin{document}

\title{Systematic effects in the measurement of polarization by the PLANCK telescope}
\author{ G.~Franco\inst{1},
         P.~Fosalba\inst{1,2},
         J.~A.~Tauber\inst{1}}

\institute{Research and Science Support Department of ESA, ESTEC, P.O.
  Box 299, NL-2200 AG Noordwijk, The Netherlands
  \and Institut d'Astrophysique de Paris, 98bis Boulevard Arago,
  F-75014 Paris, France}

\date{Received 2 October 2002 / Accepted 6 March 2003}


\abstract{We present estimates of the response to polarized signals
  by the PLANCK telescope. These estimates are based on a set of
  simulations, using a physical optics code (GRASP8), for linearly
  polarized detectors at different frequencies and located in different
  positions of an early design of the PLANCK focal plane.
  We show how the optical aberration introduced by the studied focal
  plane configuration affects absolute and relative orientation of
  the sky signals polarization planes.
  In addition, we compute the spurious signal introduced by the
  telescope optics into a PLANCK-type measurement of the sky polarization.
  Our analysis shows that the spurious polarization expected in a
  PLANCK-like experiment is typically of the order of $0.2\%$ of
  the unpolarized intensity. }
\titlerunning{Polarization systematics of the PLANCK telescope}
\authorrunning{G. Franco, P. Fosalba \& J.A. Tauber \qquad }
\maketitle


\section{Introduction}
\label{sec:intro}

The CMB temperature anisotropy carries an essential information on the
origin and evolution of large scale structures in the universe.  Its
accurate measurement by experiments probing large (see e.g.,
\cite{Bennett94}) and small scales (see e.g., \cite{Scott02} and
references therein) already allows us to constrain cosmological models
with good precision (see e.g., \cite{Lewis02}).
In addition, the quadrupole temperature anisotropy at the last
scattering surface generates a polarization signal (of order 10\% of
the CMB intensity) carried by the scalar and tensor modes of the CMB
anisotropy (\cite{Rees68}). This polarized anisotropy is a unique probe
of primordial gravitational waves (\cite{Crittenden93}) and the
reionization era of the universe (see e.g., \cite{Zaldarriaga97a};
\cite{Hu00}).  Moreover, CMB polarization helps break the
degeneracy affecting the cosmological parameter estimation from CMB
anisotropy data alone (e.g., \cite{Zaldarriaga97b}).

Detecting CMB polarization is a serious experimental challenge which
requires an unprecedented experimental sensitivity, good knowledge
and clean removal of foreground contamination, and a careful analysis
of systematic effects, the latter recently attracting increasing
attention (see e.g., \cite{Carretti01}; \cite{Kaplan01}; \cite{Leahy01};
\cite{yurchenko01}; \cite{Fosalba02}). Current upper limits for
CMB polarizations are at the level of $10~\mu$K (\cite{Hedman01};
\cite{Keating01}; \cite{Oliveira02}).
Despite great experimental efforts, CMB polarization
has remained undetected until very recently with the first detection of
$E$-mode polarization by the DASI experiment (\cite{Kovac2002}).
A new generation of CMB experiments is being specifically designed
to image the polarization anisotropies in the microwave sky with
ground-based telescopes (e.g., COMPASS), interferometers (e.g., CBI),
balloons (MAXIPOL, B2K), and satellites (MAP, PLANCK).
In particular, the PLANCK experiment\footnote{PLANCK is the
  third medium-sized mission in the Horizons 2000 Scientific Program
  of the European Space Agency, {\tt http://astro.estec.esa.nl/Planck}.}
will provide multi-frequency full-sky maps of the polarized sky with a
sensitivity better than $5\mu$K for practically all the channels
(\cite{Tauber00}).
Ultimately, the polarization performance of PLANCK depends on the
ability to minimize spurious instrumental effects which contaminate
the sky signal.

In this paper, we investigate the impact of one specific instrumental
systematic - the optical aberration in the PLANCK telescope - in the
measurement of polarized signals. The PLANCK telescope is an off-axis
aplanatic system with a very wide field of view of $\sim 8^{\circ}$ on
the sky. The focal plane is filled with individual corrugated horns
that collect radiation from the telescope and deliver it to
band-limited detectors (with bandwidth $\sim25\%$) within the 30-850
GHz frequency range. Detectors collect largely linearly polarized
radiation, defined either by a waveguide hybrid discriminator or by a
linear grid.  This experimental scheme is common to other current CMB
polarization experiments (e.g., MAP).

Although PLANCK was not originally devised as a CMB polarization
imager, its current design incorporates $7$ polarized channels between
$30$ and $353$ GHz, in the Low- and High-frequency instruments (LFI
and HFI hereafter), including a total of $88$ linearly polarized
detectors.  PLANCK, to be launched in 2007, is currently in its Phase
B (detailed design) during which many features, such as the
focal plane layout, are being optimised for the polarization
measurement. The scheme exploits the fact that many detectors in
the focal plane are sensitive to the same frequency band and are
spatially arranged such that their polarization planes are rotated
with respect to each other by multiples of $45^{\circ}$.
In this way, by combining samples from several linearly polarized
detectors, it is possible to extract all the polarization information
in terms of the Stokes parameters (see e.g., \cite{Kraus82};
\cite{Couchot99}).
However, an optimal measurement relies critically on the similarity
of the polarized angular responses, also called {\it beam} or
{\it radiation patterns}, of the differently oriented polarized horns.
These radiation patterns will be difficult to measure accurately on
the ground and they will have to be reconstructed in flight using
celestial sources such as planets and the Galactic plane.

In this paper we shall study two basic issues regarding the ability
of PLANCK to measure polarization:

\begin{itemize}

\item What is the effect of the feed and telescope optics on the
  properties of the measured polarized sky signal?
  In particular, are the absolute and relative orientation of
  polarization planes in the sky preserved by the optics?

\item What is the impact of the instrumental optics
  on the differential measurement of polarization?

\end{itemize}

In what follows we shall use Physical Optics (PO) modeling to
simulate the interaction of radiation with PLANCK telescope
optical system. Unlike Geometrical optics (GO), which is
characterized by the neglect of the wavelength, PO uses the
concepts of electromagnetic transverse waves and their
propagation properties to simulate the behaviour of light
in optical systems, being the most accurate method to
simulate diffraction, interference and polarization phenomena.
In this study we use GRASP8, a software package developed
by \cite{Ticra97}, to compute the polarized radiation pattern
on the sky for different transmitting feeds located in the
focal plane\footnote {By the principle of reciprocity, modeling
  a feed transmitting from the focal plane to the sky is
  equivalent to modeling a detector horn, in the same
  position, collecting radiation from the sky, since,
  by time reversal, the {\em transmission} pattern is
  the same as the {\em reception} pattern.}
(see \S\ref{subsec:input} for further details). In particular,
we shall simulate the feeds located at three representative (off-axis)
positions and at two different frequencies, and examine the effect of
the optical asymmetries on the measurement of the Stokes parameters.
Using such simulations we aim at answering the two basic questions
mentioned above and draw some general conclusions regarding the
impact of the PLANCK design under study in this paper, for the
measurement of CMB polarization.

The paper is organized as follows: in \S \ref{sec:theory} we address
the problem of simulating and comparing the beam (or radiation)
patterns generated using PO modeling. Results of the simulations
are presented and discussed in \S \ref{sec:results} and our final
conclusions are given in \S \ref{sec:conclusions}. In the Appendix
\ref{sec:simul} we give a full description of the simulations
procedures with GRASP8 code.

\section{Modeling Far-field Radiation Patterns}
\label{sec:theory}

\subsection{Introducing reference frames}
\label{subsec:frames}

Polarized radiation can be fully described by the projection of the
electric field vector, ${\bf E}$, in a 2-dimensional orthogonal
basis. For the far field radiation pattern, ${\bf E_{far}}$, we shall adopt Ludwig's 3rd definition
of co- and cross-polarization (\cite{Ludwig73}),
\begin{eqnarray}
{\bf E_{far}} &=& E_{co}\sigma_{co}+E_{cr}\sigma_{cr}
\label{eq:Efar}
\end{eqnarray}
where $\sigma_{co}$ and $\sigma_{cr}$ are the co- and
cross-polarization unit vectors which relate to the usual spherical
polar basis ($\sigma_{\theta}$,~$\sigma_{\phi}$) as follows,
\begin{eqnarray}
\sigma_{co} &=& \sigma_{\theta}\sin\phi+\sigma_{\phi}\cos\phi \nonumber\\
\sigma_{cr} &=& \sigma_{\theta}\cos\phi-\sigma_{\phi}\sin\phi .
\label{eq:unitco}
\end{eqnarray}
By projecting onto the Cartesian coordinate system,
\begin{eqnarray}
\sigma_{\theta} &=& \sigma_x\cos\theta\cos\phi+\sigma_y\cos\theta\sin\phi-\sigma_z\sin\theta \nonumber\\
\sigma_{\phi} &=& -\sigma_x\sin\phi+\sigma_y\cos\phi
\label{eq:unitheta}
\end{eqnarray}
one finds that at boresight, i.e., the north pole of the sphere
($\theta\rightarrow0^{\circ}$), it holds that
$\sigma_{co}\rightarrow \sigma_y$ and $\sigma_{cr}\rightarrow\sigma_x$
(see eqs. (\ref{eq:unitco}) and (\ref{eq:unitheta})),
so the chosen polarization basis lies along the Cartesian one.
Away from boresight, Ludwig's polarization basis is obtained by
parallel-transporting the local Cartesian basis ($\sigma_x$,~$\sigma_y$)
along great circles over the sphere.

\subsection{Defining cross-polarization for non-ideal antennas}
\label{subsec:xpd}

Here we shall define a robust way of characterizing cross-polarization
in non-ideal optical systems. In particular, we discuss cross-polarization
definitions for the case of feeds with arbitrary orientations in the focal
plane.

Let us denote the orientation of a given feed (with respect to its
symmetry or rotation axis) by $\psi_{FP}$.  We define a reference
orientation, denoted by $\psi_{FP}^{nr}=0^{\circ}$, with respect
to which we can define any arbitrary orientation of a given feed.
Similarly, we must introduce a corresponding far-field reference
frame with respect to which the sky beam patterns from a given feed
with an arbitrary orientation in the focal plane, will be referred to.

For an ideal telescope\footnote{We define an ideal telescope as one
  with rotationally symmetric reflectors and on-axis fully linearly
  polarized detectors.} with an on-axis feed, one can always find a
reference frame in the far field for which the cross-polarization power
(see~eq.(\ref{eq:Efar})) vanishes ($E^2_{cr}=0$). This constitutes
a convenient reference frame for sky beam patterns of a given feed.
However, for a real telescope, any asymmetry in the antenna system
will generally introduce a non-vanishing cross-polarization component
in the far-field radiation pattern.  Moreover, non-ideal telescope
optics also introduce a {\it differential rotation} of the principal
plane of polarization in the far field with respect to that physically
applied to the feed in the focal plane.

We recall that a proper measurement of polarization in terms of
the Stokes parameters of the sky signal requires that the orientation
of the feeds in the focal plane be such that the principal planes of
polarization {\it on the sky} are spaced by $45^{\circ}$.
Therefore, it is important to choose an appropriate method for
determining this principal plane of polarization, which is here
defined as that in which the Cross Polarization Discrimination
(XPD) is maximized.

We define {\it maximal cross-polar discrimination angle}, $\phi_{XPD}$,
as the angle by which one should rotate the reference frame in the
sky in order to maximize the Cross Polar Discrimination, $XPD$.
Below we introduce three different methods for the computation
of $\phi_{XPD}$, each one relying on a different definition of
XPD\footnote{For simplicity  of notation, XPD will always be
  expressed in dB units as a power difference, $\Delta W$.}:

\begin{itemize}
\item The {\bf 1st Method} is the most commonly used in the literature.
  $\phi_{XPD}$ is the angle that maximizes XPD defined as the ratio
  of the (single) co-polar power peak to the {\it highest}
  cross-polarization power peak. In dB units we can write the
  maximized XPD as,
  \begin{equation}
   \Delta W_{max}=\Delta W (\phi_{XPD}) =  10\,\log_{10}\left[\frac
   {Max(E^{rot}_{co}E^{rot*}_{co})}{Max(E^{rot}_{cr}E^{rot*}_{cr})}
   \right]_{(\phi=\phi_{XPD})}
   \!\!\!\!\!\!\!\!\!\!\!\!\!\!\!\!\!\!\!\!\!\!=\,\,
   10\,\log_{10}\,[Max(\,(E^{rot}_{co})^2\,)]-
   10\,\log_{10}\,[Max(\,(E^{rot}_{cr})^2\,)]
   \label{eq:maxdifpk}
  \end{equation}
\item In the {\bf 2nd Method}, XPD is defined as the difference between
  the co-polar power peak and the cross-polarization power
  {\it along the same direction}. Since $\phi_{XPD}$ is the angle that
  maximizes this difference, we can write
  \begin{equation}
   \Delta W_{max}=\Delta W (\phi_{XPD}) = 10\,\log_{10}\,
   [Max(\,(E^{rot}_{co})^2\,)]-10\,\log_{10}\,[(E^{rot}_{cr}(\theta_{Max},
   \phi_{Max})\,)^2]
   \label{eq:maxdifcp}
  \end{equation}
  where $\theta_{Max}$ and $\phi_{Max}$ are the coordinates of the
  co-polar power peak in the spherical coordinate system
  $(\theta,\phi)$.
\item The {\bf 3rd Method} determines $\phi_{XPD}$ as the angle that
  maximizes the difference between the {\it integrated} co-polar and
  cross-polarization amplitudes over the main beam, and thus the
  maximized XPD is given by
  \begin{equation}
   \Delta W_{max}=\Delta W (\phi_{XPD}) =
   10\,\log_{10}\left[\int_{-\theta_0}^{+\theta_0}\int_0^{\pi}
   (E^{rot}_{co})^2\,d\Omega\right] -
   10\,\log_{10}\left[\int_{-\theta_0}^{+\theta_0}\int_0^{\pi}
   (E^{rot}_{cr})^2\,d\Omega\right]
\label{eq:maxdifint}
  \end{equation}
  where the integration interval on $\theta$,
  $[ -\theta_0 ; +\theta_0 ]$, must be taken wide enough to
  encompass most of the power in the co- and cross-polarization
  patterns. We chose $\theta_0$ to be at least twice the
  Full Width at Half Maximum (FWHM) of the co-polar power pattern.
\end{itemize}

Note that in the equations above $E^{rot}_{co}$ and $E^{rot}_{cr}$
are the (complex) co- and cross-polarization components of the
radiation pattern, after having performed a rotation of the far
field coordinate system by an angle ($\phi_{XPD}$ in this case)
with respect to the original reference frame (i.e, the far-field
frame defined for a non-rotated feed in the focal plane),
\begin{eqnarray}
E_{co}^{rot} &=& E_{cr}\sin\phi_{XPD}+E_{co}\cos\phi_{XPD}\nonumber\\
E_{cr}^{rot} &=& E_{cr}\cos\phi_{XPD}-E_{co}\sin\phi_{XPD}
\label{eq:Erot}
\end{eqnarray}

In general, the far-field radiation pattern of an asymmetric radiating
system fed by a corrugated horn has a cross-polarization pattern with
a {\it multiple} peak (rather than a simple peak) structure with
several maxima which are not spatially coincident with nor
symmetrically placed with respect to the co-polar peak
(e.g. see Figure~\ref{fig:inpattern}). Such an asymmetric
cross-polarization pattern for non-ideal antennas makes the cross
polar discrimination angle defined in the 1st method potentially
ambiguous. In principle, the 3rd method should be appropriate for
measurements of diffuse polarized emission, as it is based on
integrating radiation filling the main beam (i.e, an extended
region around boresight), while the 2nd  method is more appropriate
for measuring polarization from point sources as it relies on a
local measurement of power radiated.

\subsection{Differential rotation of the polarization plane}
\label{subsec:rotation}

Let us consider again an ideal antenna system. In this case, it is
expected that a rotation of a given feed around its own symmetry axis,
from an initial position, $\psi_{FP}^i$, to a final position,
$\psi_{FP}^f$, would result in an equal rotation of the principal
plane of polarization of the radiation pattern in the sky\footnote{In
  fact, in a dual reflector system such as that of PLANCK, the rotation
  of the principal plane of polarization on the sky would be
  {\em symmetric} and not {\em equal} to the rotation of the feed, due
  to the ``mirroring'' of the radiation pattern in the beam coordinate
  system when the signal is reflected from one mirror to the other.
  Equation (\ref{eq:ideal}) takes this symmetry into account.}. Thus we
can write,
\begin{equation}
\phi_{XPD}^f-\phi_{XPD}^i=-(\psi_{FP}^f-\psi_{FP}^i)
\label{eq:ideal}
\end{equation}
Setting $\phi_{XPD}^i=\phi_{XPD}^{nr}$ as the maximal
cross-polarization discrimination angle for the beam pattern
when the feed is in its original or non-rotated position
($\psi_{FP}^i=0^{\circ}$), the previous equation becomes,
\begin{equation}
\phi_{XPD}-\phi_{XPD}^{nr}=-\psi_{FP} \, .
\label{eq:equal}
\end{equation}
This equation holds for any ideal antenna system if the polarization
properties of the radiation pattern are equally preserved for all
orientations of the feed in the focal plane.

However, for a non-ideal antenna system, such isotropy is broken and,
due to factors such as the asymmetry of the reflectors or the
off-axis positioning of the feed, equation (\ref{eq:equal}) will
only hold to a first-order approximation, i.e.,
$\phi_{XPD}-\phi_{XPD}^{nr}\approx-\psi_{FP}$.
Consequently, in order to quantify how good this approximation is (or
on the contrary, how non-ideal is the antenna), we shall define the
{\em residual angle}, $\Delta_{\phi}$, as the additional angle to
which the reference frame of the beam pattern in the far-field should
be rotated, in addition to $(-\psi_{FP})$, so as to have the co-polar
axis of the beam pattern reference frame aligned with the principal
polarization direction.  Thus we define
\begin{equation}
\Delta_{\phi}\equiv\phi_{XPD}-\phi_{XPD}^{nr}-(-\psi_{FP})
\label{eq:delta}
\end{equation}
where $\phi_{XPD}$ is computed using the methods described in
\S\ref{subsec:xpd} and $\phi_{XPD}^{nr}$ is the angle $\phi_{XPD}$ for
the non-rotated feed, $\phi_{XPD}^{nr} = \phi_{XPD}(\psi_{FP}=0^{\circ})$.

\section{Simulation Results}
\label{sec:results}

In what follows, we shall focus on determining this {\em residual
angle}, along with the associated angle $\phi_{XPD}$ and power
difference $\Delta W (\phi_{XPD})$, making use of the PO
simulations described in \S\ref{subsec:input}.
In the present study we compute the sky beam patterns for
PLANCK-LFI 30 and 100~GHz feeds at different positions in
the focal plane unit, for a range of different orientations
of the feed.  The orientations studied correspond to rotations
around the symmetry axis of the feed by angles\footnote{Note that,
  by symmetry, a feed rotation by $\psi_{FP}$ is equivalent to a
  $\psi_{FP} - 180^{\circ}$ rotation, so we need not consider
  cases with $\psi_{FP}>180^{\circ}$.} $\psi_{FP} = 0^{\circ},
45^{\circ}, 60^{\circ}, 90^{\circ}, 120^{\circ}, 135^{\circ}$
and $180^{\circ}$.
In particular, we carry out numerical simulations of the far-field
patterns fed by corrugated horns located at three different positions
in the focal plane ranging from very close to the optical axis, to the
edge of the field of view (see Figure~\ref{fig:focalplane}).

This will allow us to draw some general conclusions about the effect
of the feed positioning on the preservation of the polarization
properties through the antenna.  We emphasize that, despite
concentrating on the PLANCK experiment, our results can be useful for
other mm/cm-wave polarimetric experiments with similar optical
systems.

\subsection{Effect of feed position and orientation}
\label{subsec:feedrotation}

Tables \ref{tab:resultspk}, \ref{tab:resultscp} and \ref{tab:resultsint}
display the values of the {\em residual angle} ($\Delta_{\phi}$) for all
the simulated cases. These values were computed using equation
(\ref{eq:delta}).
The corresponding values for $\phi_{XPD}$ and $\Delta W_{max}$ were
determined using the three different methods described in
\S\ref{subsec:xpd}. In each table, each $\Delta_{\phi}$-column
shows the results for a different location of the feed in the
focal plane (see Figure~\ref{fig:focalplane}).
We point out that, when using the 3rd Method, the main beam
pattern is integrated over a solid angle centered at boresight
(the north pole of the sphere) and down to a latitude
$\theta_0=1.0^{\circ}$ and $0.5^{\circ}$ for the 30 and 100~GHz
feeds, respectively. These solid angles are chosen so as to include,
at least, all radiated power down to the FWHM (see \S\ref{subsec:xpd}).


\begin{table*}[t!]
\small
\centerline{\begin{tabular}{|c c c c c c c c c c|}
\multicolumn{10}{c}{\bf Results on the Determination of $\Delta_{\phi}$
by Maximizing the Difference}\\
\multicolumn{10}{c}{\bf Between the Co- and Cross-polar Power Peaks
(1st Method)}\\
\hline\hline
{\bf30 GHz} &
\multicolumn{3}{c|}{Position 1} &
\multicolumn{3}{c|}{Position 4} &
\multicolumn{3}{c|}{Position 27} \\
\hline
$\psi_{FP}$&
$\phi_{XPD}$&$\Delta W_{Max}$&$\Delta_{\phi}$&
$\phi_{XPD}$&$\Delta W_{Max}$&$\Delta_{\phi}$&
$\phi_{XPD}$&$\Delta W_{Max}$&$\Delta_{\phi}$\\
$(^{\circ})$&
$(^{\circ})$&$(dB)$&$(^{\circ})$&
$(^{\circ})$&$(dB)$&$(^{\circ})$&
$(^{\circ})$&$(dB)$&$(^{\circ})$\\
\hline
0   &    0.4 & 33 &    0 &    4.8 & 29 &   0 &    7.9 & 26 &   0 \\\hline
45  &  -43.9 & 33 &  0.7 &  -39.1 & 29 & 1.2 &  -35.1 & 26 &   2 \\\hline
60  &  -58.9 & 33 &  0.7 &  -54.0 & 29 & 1.2 &  -49.7 & 26 & 2.4 \\\hline
90  &  -89.3 & 33 &  0.3 &  -83.8 & 29 & 1.4 &  -79.6 & 26 & 2.5 \\\hline
120 & -119.9 & 33 & -0.3 & -114.2 & 29 & 1.0 & -110.8 & 26 & 1.3 \\\hline
135 & -135.0 & 33 & -0.4 & -129.8 & 29 & 0.5 & -126.5 & 26 & 0.6 \\\hline
180 & -179.6 & 33 &    0 & -175.2 & 29 &   0 & -172.1 & 26 &   0 \\\hline
\multicolumn{10}{c}{}\\
\hline\hline
{\bf100 GHz} &
\multicolumn{3}{c|}{Position 1} &
\multicolumn{3}{c|}{Position 4} &
\multicolumn{3}{c|}{Position 27} \\
\hline
$\psi_{FP}$&
$\phi_{XPD}$&$\Delta W_{Max}$&$\Delta_{\phi}$&
$\phi_{XPD}$&$\Delta W_{Max}$&$\Delta_{\phi}$&
$\phi_{XPD}$&$\Delta W_{Max}$&$\Delta_{\phi}$\\
$(^{\circ})$&
$(^{\circ})$&$(dB)$&$(^{\circ})$&
$(^{\circ})$&$(dB)$&$(^{\circ})$&
$(^{\circ})$&$(dB)$&$(^{\circ})$\\
\hline
0   &    0.3 & 32 &    0 &    5.9 & 30 &    0 &    9.3 & 28 &  0.0 \\\hline
45  &  -43.8 & 32 &  0.9 &  -39.9 & 30 & -0.8 &  -36.1 & 28 & -0.4 \\\hline
60  &  -58.8 & 32 &  0.9 &  -55.1 & 30 & -1.0 &  -51.3 & 28 & -0.6 \\\hline
90  &  -89.4 & 31 &  0.3 &  -85.0 & 30 & -0.9 &  -80.4 & 27 &  0.3 \\\hline
120 & -120.0 & 32 & -0.3 & -114.8 & 30 & -0.7 & -110.4 & 27 &  0.3 \\\hline
135 & -135.3 & 32 & -0.6 & -129.5 & 30 & -0.4 & -125.6 & 27 &  0.1 \\\hline
180 & -179.7 & 32 &    0 & -174.1 & 30 &    0 & -170.7 & 28 &    0 \\\hline
\end{tabular}}
\caption[]{\small{Maximal cross-polarization discrimination angle
  $\phi_{XPD}$,  $\Delta W_{max}$ and the {\em residual angle}
  $\Delta_{\phi}$ for the {\bf 30 GHz} and the {\bf 100 GHz} feeds
  on positions 1, 4 and 27, for different orientations $\psi_{FP}$
  of the feeds being $\phi_{XPD}$ determined by the {\bf 1st Method}
  described in \S\ref{subsec:xpd} and $\Delta W_{Max}$ evaluated
  using equation (\ref{eq:maxdifpk}).}}
\label{tab:resultspk}
\end{table*}


\begin{table*}[t!]
\small
\centerline{\begin{tabular}{|c c c c c c c c c c|}
\multicolumn{10}{c}{\bf Results on the Determination of $\Delta_{\phi}$
by Maximizing the Difference Between}\\
\multicolumn{10}{c}{\bf the Co-polar Power Peak and the Cross-polar
Component (2nd Method)}\\
\hline\hline
{\bf30 GHz} &
\multicolumn{3}{c|}{Position 1} &
\multicolumn{3}{c|}{Position 4} &
\multicolumn{3}{c|}{Position 27} \\
\hline
$\psi_{FP}$&
$\phi_{XPD}$&$\Delta W_{Max}$&$\Delta_{\phi}$&
$\phi_{XPD}$&$\Delta W_{Max}$&$\Delta_{\phi}$&
$\phi_{XPD}$&$\Delta W_{Max}$&$\Delta_{\phi}$\\
$(^{\circ})$&
$(^{\circ})$&$(dB)$&$(^{\circ})$&
$(^{\circ})$&$(dB)$&$(^{\circ})$&
$(^{\circ})$&$(dB)$&$(^{\circ})$\\
\hline
0   &    0.5 & 60 &    0 &    5.7 & 56 &    0 &    8.9 & 43 &    0 \\\hline
45  &  -44.4 & 60 &  0.1 &  -39.4 & 65 & -0.1 &  -36.4 & 51 & -0.3 \\\hline
60  &  -59.4 & 60 &  0.1 &  -54.5 & 61 & -0.2 &  -51.5 & 53 & -0.4 \\\hline
90  &  -89.5 & 60 &    0 &  -84.5 & 76 & -0.2 &  -81.5 & 50 & -0.4 \\\hline
120 & -119.5 & 61 &    0 & -114.4 & 57 & -0.1 & -111.4 & 44 & -0.3 \\\hline
135 & -134.5 & 61 &    0 & -129.4 & 54 & -0.1 & -126.3 & 43 & -0.2 \\\hline
180 & -179.5 & 60 &    0 & -174.3 & 56 &    0 & -171.1 & 43 &    0 \\\hline
\multicolumn{10}{c}{}\\
\hline\hline
{\bf100 GHz} &
\multicolumn{3}{c|}{Position 1} &
\multicolumn{3}{c|}{Position 4} &
\multicolumn{3}{c|}{Position 27} \\
\hline
$\psi_{FP}$&
$\phi_{XPD}$&$\Delta W_{Max}$&$\Delta_{\phi}$&
$\phi_{XPD}$&$\Delta W_{Max}$&$\Delta_{\phi}$&
$\phi_{XPD}$&$\Delta W_{Max}$&$\Delta_{\phi}$\\
$(^{\circ})$&
$(^{\circ})$&$(dB)$&$(^{\circ})$&
$(^{\circ})$&$(dB)$&$(^{\circ})$&
$(^{\circ})$&$(dB)$&$(^{\circ})$\\
\hline
0   &    0.5 & 64 &    0 &    6.1 & 56 &    0 &    9.2 & 38 &    0 \\\hline
45  &  -44.4 & 58 &  0.1 &  -39.5 & 61 & -0.6 &  -36.8 & 38 & -1.0 \\\hline
60  &  -59.4 & 57 &  0.1 &  -54.7 & 58 & -0.8 &  -52.1 & 38 & -1.3 \\\hline
90  &  -89.5 & 61 &    0 &  -84.8 & 59 & -0.9 &  -82.2 & 37 & -1.4 \\\hline
120 & -119.5 & 69 &    0 & -114.4 & 66 & -0.5 & -111.5 & 37 & -0.7 \\\hline
135 & -134.5 & 62 &    0 & -129.2 & 61 & -0.3 & -126.1 & 37 & -0.3 \\\hline
180 & -179.5 & 64 &    0 & -173.9 & 56 &    0 & -170.8 & 38 &    0 \\\hline
\end{tabular}}
\caption[]{\small{Same as Table \ref{tab:resultspk}, with $\phi_{XPD}$ being
   determined by the {\bf 2nd Method}, as described in \S\ref{subsec:xpd}.}}
\label{tab:resultscp}
\end{table*}


\begin{table*}[t!]
\small
\centerline{\begin{tabular}{|c c c c c c c c c c|}
\multicolumn{10}{c}{\bf Results on the determination of $\Delta_{\phi}$
by Maximizing the Difference}\\
\multicolumn{10}{c}{\bf Between the Integrated Co- and Cross-polar
Power Beams (3rd Method)}\\
\hline\hline
{\bf30 GHz} &
\multicolumn{3}{c|}{Position 1} &
\multicolumn{3}{c|}{Position 4} &
\multicolumn{3}{c|}{Position 27} \\
\hline
$\psi_{FP}$&
$\phi_{XPD}$&$\Delta W_{Max}$&$\Delta_{\phi}$&
$\phi_{XPD}$&$\Delta W_{Max}$&$\Delta_{\phi}$&
$\phi_{XPD}$&$\Delta W_{Max}$&$\Delta_{\phi}$\\
$(^{\circ})$&
$(^{\circ})$&$(dB)$&$(^{\circ})$&
$(^{\circ})$&$(dB)$&$(^{\circ})$&
$(^{\circ})$&$(dB)$&$(^{\circ})$\\
\hline
0  &   0.5&$32$& 0 &   5.6&$28$& 0 &   8.7&$25$& 0\\\hline
45 & -44.5&$32$& 0 & -39.4&$28$& 0 & -36.3&$25$& 0\\\hline
60 & -59.5&$32$& 0 & -54.4&$28$& 0 & -51.3&$25$& 0\\\hline
90 & -89.5&$32$& 0 & -84.4&$28$& 0 & -81.3&$25$& 0\\\hline
120&-119.5&$32$& 0 &-114.4&$28$& 0 &-111.3&$25$& 0\\\hline
135&-134.5&$32$& 0 &-129.4&$28$& 0 &-126.3&$25$& 0\\\hline
180&-179.5&$32$& 0 &-174.4&$28$& 0 &-171.3&$25$& 0\\\hline
\multicolumn{10}{c}{}\\
\hline\hline
{\bf100 GHz} &
\multicolumn{3}{c|}{Position 1} &
\multicolumn{3}{c|}{Position 4} &
\multicolumn{3}{c|}{Position 27} \\
\hline
$\psi_{FP}$&
$\phi_{XPD}$&$\Delta W_{Max}$&$\Delta_{\phi}$&
$\phi_{XPD}$&$\Delta W_{Max}$&$\Delta_{\phi}$&
$\phi_{XPD}$&$\Delta W_{Max}$&$\Delta_{\phi}$\\
$(^{\circ})$&
$(^{\circ})$&$(dB)$&$(^{\circ})$&
$(^{\circ})$&$(dB)$&$(^{\circ})$&
$(^{\circ})$&$(dB)$&$(^{\circ})$\\
\hline
0  &   0.5 &$33$& 0 &   5.5 &$29$&   0 &   8.7 &$26$&   0\\\hline
45 & -44.5 &$33$& 0 & -39.4 &$29$& 0.1 & -36.2 &$26$& 0.1\\\hline
60 & -59.5 &$33$& 0 & -54.4 &$29$& 0.1 & -51.2 &$26$& 0.1\\\hline
90 & -89.5 &$33$& 0 & -84.4 &$29$& 0.1 & -81.2 &$26$& 0.1\\\hline
120&-119.5 &$33$& 0 &-114.5 &$29$&   0 &-111.3 &$26$&   0\\\hline
135&-134.5 &$33$& 0 &-129.5 &$29$&   0 &-126.3 &$26$&   0\\\hline
180&-179.5 &$33$& 0 &-174.5 &$29$&   0 &-171.3 &$26$&   0\\\hline
\end{tabular}}
\caption[]{\small{Same as Table \ref{tab:resultspk}, with $\phi_{XPD}$ being
   determined by the {\bf 3rd Method}, as described in \S\ref{subsec:xpd}.}}
\label{tab:resultsint}
\end{table*}

Our results show that the estimated differential rotation of the
polarization plane (see \S\ref{subsec:rotation}) in terms of the
residual angle, $\Delta_{\phi}$, depends strongly on the definition
adopted for maximum cross-polarization discrimination and the associated
angle, $\phi_{XPD}$.
In practice, we see from the calculations that the 3rd Method gives what
one would intuitively expect from a robust estimate of cross-polarization:
{\it there is no differential rotation of the polarization plane in any
of the cases studied} (i.e., the residual angle, $\Delta_{\phi}$ is
found to be compatible with zero in all cases, see
Table~\ref{tab:resultsint}).  In fact, the maximum value of
the differential rotation found ($\Delta_{\phi} = 0.1^{\circ}$) for
the 100~GHz feed at the off-axis positions 4 and 27 (see
Figure~\ref{fig:focalplane}), is not significant given the errors,
typically of this order, coming from the accuracy in the computation
of the beam pattern itself as well as its finite sampling in
spherical cuts.
On the other hand, Methods 1 and 2 are misleading when measuring
diffuse polarization and they indicate a significant rotation of
the polarization plane, whose magnitude varies systematically
with frequency and position in the focal plane (see
Tables~\ref{tab:resultspk} and \ref{tab:resultscp}).

In summary, we see that the 3rd Method turns out to be an appropriate
definition of cross-polarization and the related cross-polar rejection
angle for measurements of diffuse polarized emission. Furthermore,
making use of this definition, we conclude that {\it the PLANCK
  telescope does not introduce any second order (i.e. detector
  orientation dependent) spurious polarization.}

For the absolute orientation of the polarization plane
of the sky signal there is a {\it misalignment angle} introduced by
the telescope optics. Referring to Figures~\ref{fig:side} and
\ref{fig:oblique}, by misalignment angle we mean the angle, as
seen from the the xy-plane of the line-of-sight Cartesian coordinate
system of the telescope\footnote{By definition, the z-axis of the
  line-of sight coordinate system points to the direction of maximum
  power radiated (or received) by an on-axis detector.} ($C_{PM}$),
between the y-axis of the feed coordinate system (e.g. f1) and
the y-axis of the corresponding coordinate system of the sky beam
pattern (mb1 as corresponding to f1).

\begin{table*}[h!]
\centerline{\begin{tabular}{|c c c c|}
\hline\hline
Feed Position in Focal Plane & 1 & 4 & 27\\ \hline
Misalignment Angle  & $0.3^{\circ}$ & $1.9^{\circ}$ & $4.3^{\circ}$\\
\hline
\end{tabular}}
\caption[]{\small Misalignment angle between the y-axis of both
    the feed and the far-field coordinate systems, for each position
    of the feed in the focal plane. Feed positions are shown in
    Figure~\ref{fig:focalplane}.}
\label{tab:projections}
\end{table*}

Table \ref{tab:projections} shows the results for the misalignment
angles, computed for each simulated position of the feed in the
focal plane. They take into account the values of maximal
cross-polarization rejection for the non-rotated feed,
$(\psi_{FP}^{nr}=0^{\circ})$ (as determined using the 3rd Method).
It can be seen that the bias or misalignment angle of the
polarization plane increases as the feed moves from
the center (on-axis) towards the edge of the focal plane
(as we would expect), yielding an upper limit of about $5^{\circ}$.
Note that these results are independent of feed orientation and
frequency, within the errors (see Table~\ref{tab:resultsint}),
which means that they do not affect the differential measurement
of polarization for each position of the feed. In addition, this
bias can always be corrected by rearranging the orientation of
each detector in the focal plane.

\subsection{Predicting far-field beam patterns}
\label{subsec:beamshape}

\begin{figure}[p!]   
  \unitlength1.0cm
\begin{minipage}[t]{7.0cm}
  \parbox[t]{0cm}{} \put(0.7,1.0){\small (a) 1st Method}
  \put(9.5,1.0){\small (b) 3rd Method}
  \put(3.0,0.0){\psfig{file=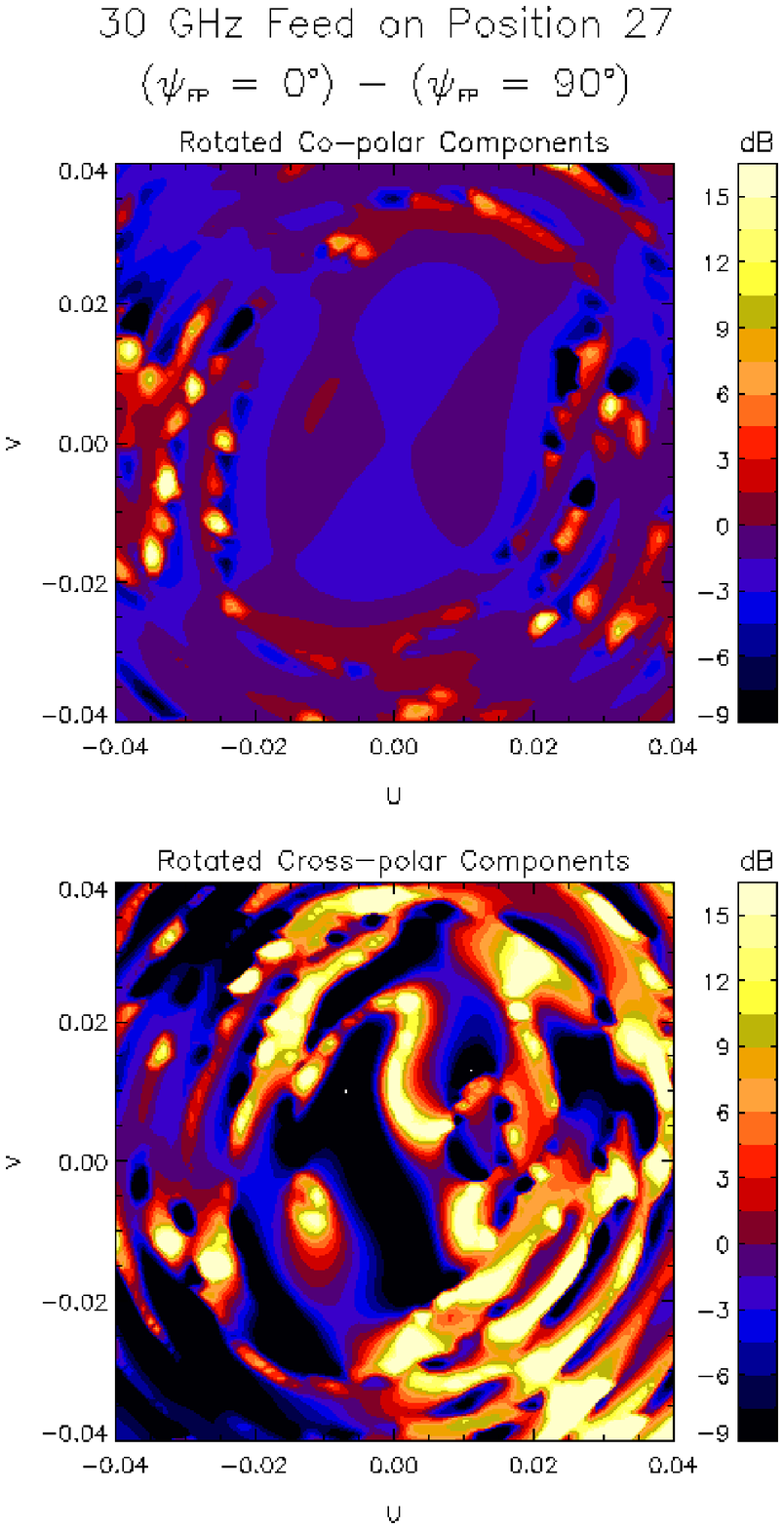,height=4.2in,clip=,silent=}}
\end{minipage} \hfill
\begin{minipage}[t]{7.0cm}
  \centerline{\psfig{file=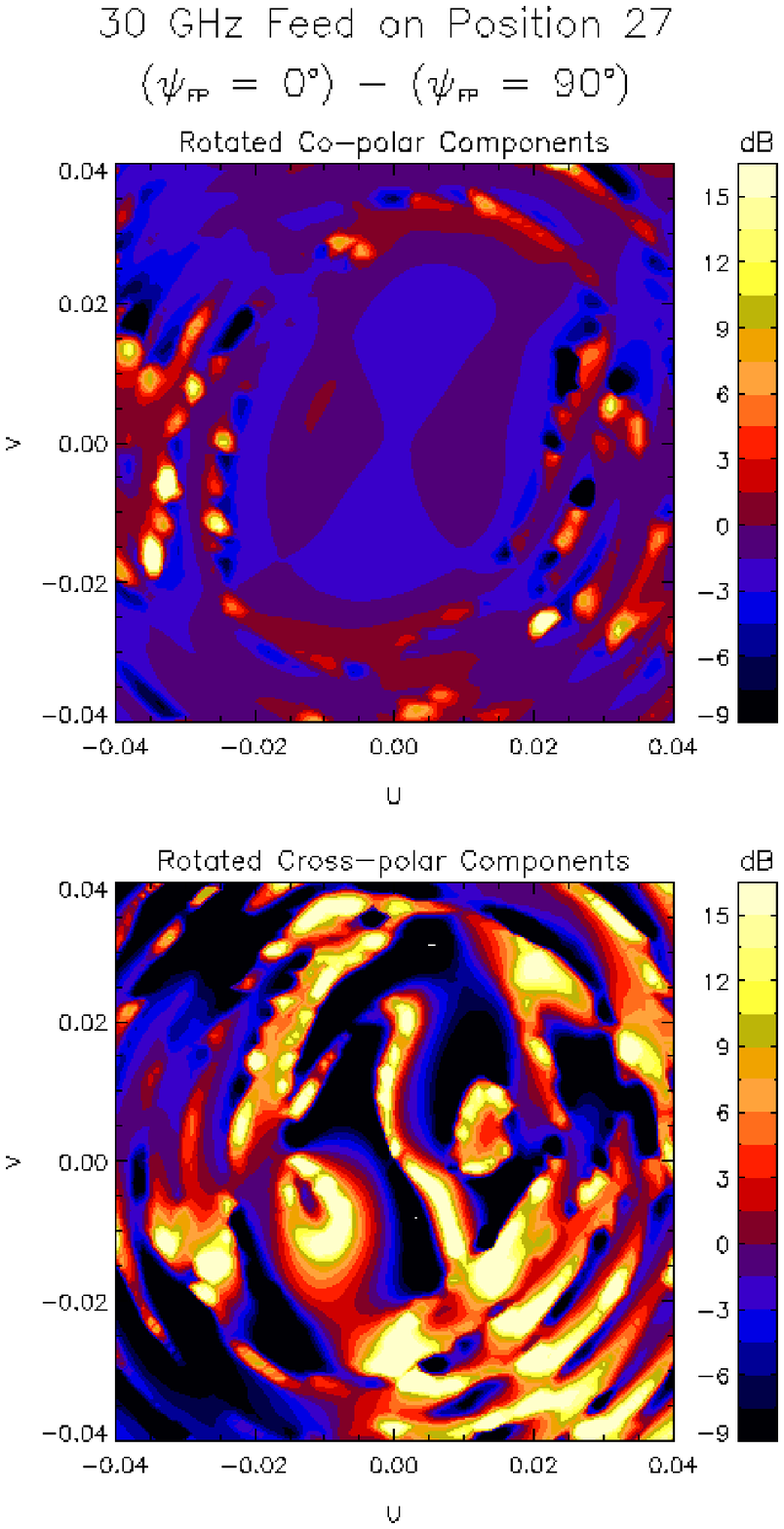,height=4.2in,clip=,silent=}}
\end{minipage}

\begin{minipage}[t]{7.0cm}
  \parbox[t]{0cm}{} \put(0.7,1.0){\small (c) 3rd Method}
  \put(9.5,1.0){\small (d) 3rd Method}
  \put(3.0,0.0){\psfig{file=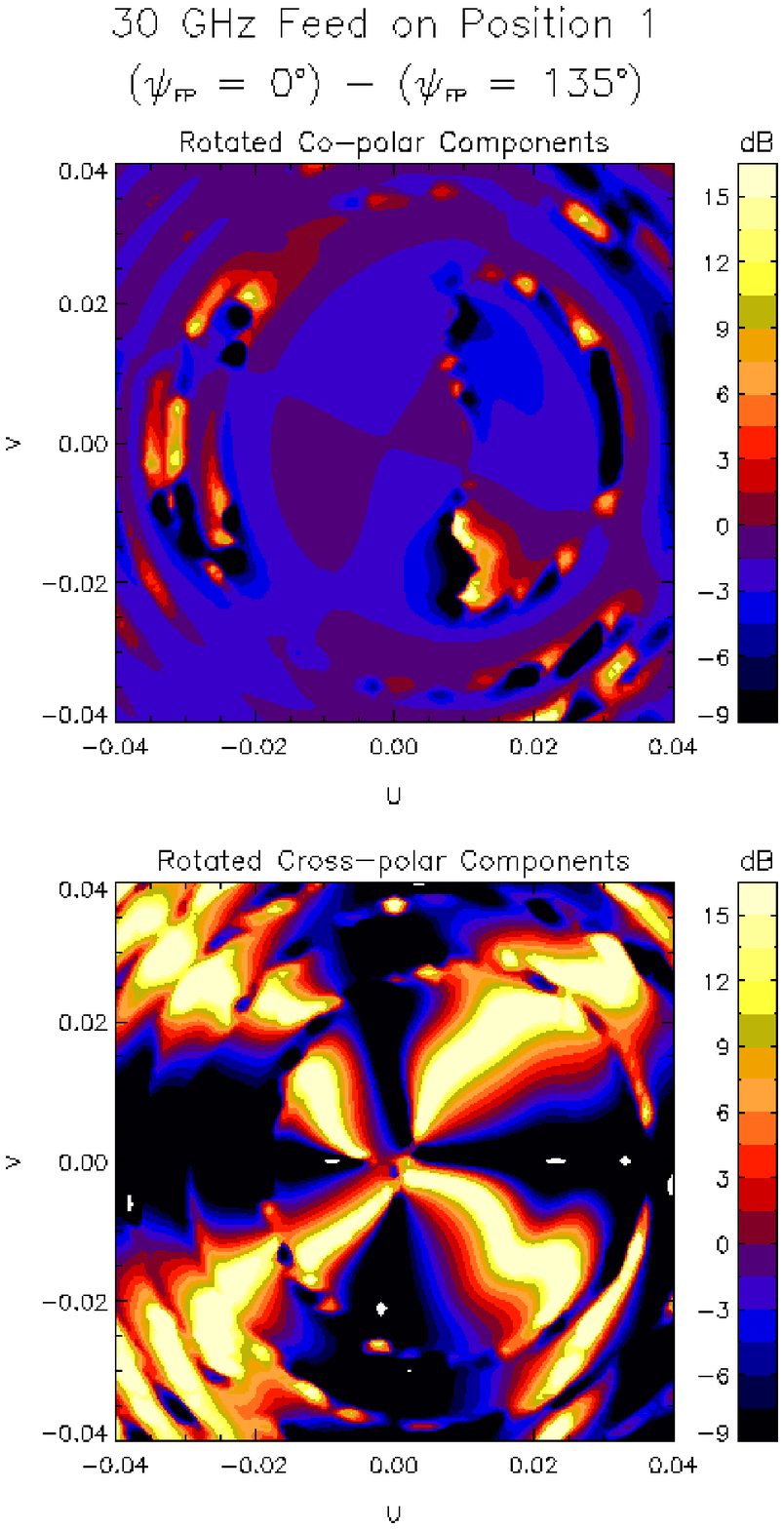,height=4.2in,clip=,silent=}}
\end{minipage} \hfill
\begin{minipage}[t]{7.0cm}
  \centerline{\psfig{file=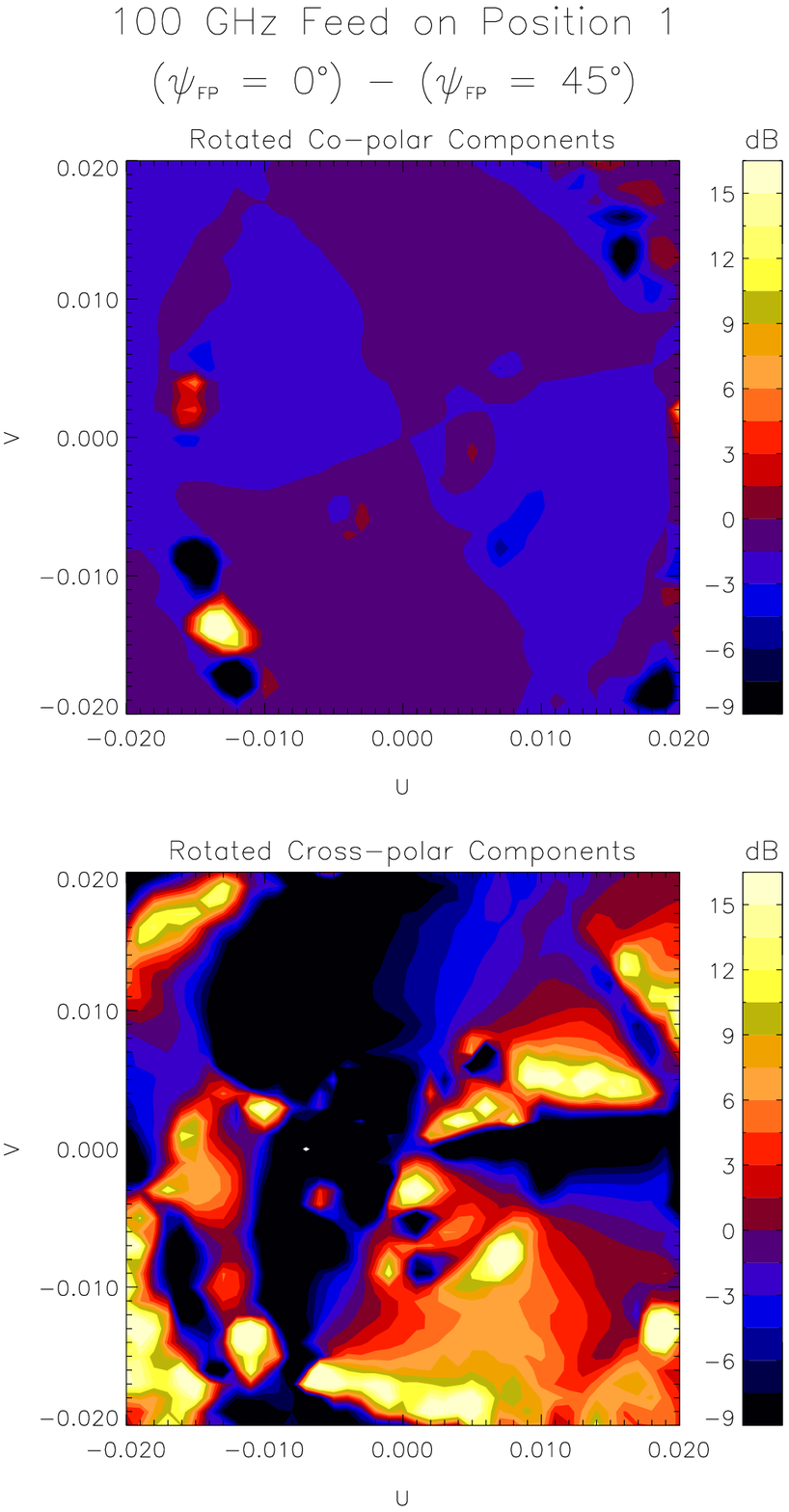,height=4.2in,clip=,silent=}}
\end{minipage}

\caption[]{Differences (in dB units) between main beam patterns in far
    field reference frames rotated by $\phi_{XPD}$, for feeds at
    different positions and orientations in the focal plane.
    Cross-polarization definitions and $\phi_{XPD}$ angles are
    determined by Methods 1~\&~3 described in \S\ref{subsec:xpd}
    (see Tables~\ref{tab:resultspk} and \ref{tab:resultsint}).}
\label{fig:contourdif}
\end{figure}

PO modeling of a large telescope such as PLANCK requires
computer intensive calculations with processing time scaling
as the fourth power of frequency. Moreover, modeling the
response of a focal plane array including many detectors
at several frequencies becomes an extremely demanding task.
Therefore, we investigate to what extent it is possible to use
a known (pre-computed) model of the main beam response (i.e, the
peak of detector angular response), for a given feed and focal
plane layout, to estimate or predict the sky main beam patterns
for that detector in different focal plane configurations, i.e.,
for the feed in the same off-axis position but different
polarization sensitivity directions.
In order to do so, we must assess the similarity between the
main beam patterns for different orientations of the
radiating feed in the focal plane.  At each frequency, we
compare the beam pattern generated from two different feed
orientations for which the far-field reference frame
has been rotated by $\phi_{XPD}$ (with its co-polar component
aligned with the principal plane of polarization).
Figure~\ref{fig:contourdif} shows the differences (in dB units)
between the power contours of several pairs of main beam patterns.
For illustrative purposes, we display the difference contours for
the case where the definitions of maximum cross-polarization
discrimination discussed in \S\ref{subsec:xpd} disagree the most
(Figures~\ref{fig:contourdif}(a),\ref{fig:contourdif}(b)) and
for the cases where the feeds are closest to the center of the focal
plane (Figures~\ref{fig:contourdif}(c),\ref{fig:contourdif}(d)).

As can be seen from Figure~\ref{fig:contourdif}, co-polar power
differences are always below $3.0$~dB, except for a few point-like
regions, while cross-polar power mismatches are locally found up to
$15$~dB. Since cross-polarization peaks are typically about $30$~dB
bellow the co-polar peak (see $\Delta W_{Max}$ values in
Table~\ref{tab:resultsint}), it is sufficient to concentrate
on the dominant (co-polar) component of the polarized beam
pattern to estimate how much the beam pattern changes as a
function of the feed orientation.
No significant difference is found between the co-polar patterns
on Figures~\ref{fig:contourdif}(a) and \ref{fig:contourdif}(b)
and therefore we conclude that these difference contours cannot
be used to determine a robust definition for cross-polarization.

More interestingly, from Figure \ref{fig:contourdif},
we conclude that using a pre-computed sky main beam pattern
from a feed at a given frequency and orientation in the focal plane
(and the corresponding maximal cross-polarization rejection angle,
$\phi_{XPD}$), it is possible to predict the shape of the main
beam pattern in the sky for any other orientation of the same feed,
within $\sim$3~dB ($\sim$15~dB) for the the co-polar
(cross-polar) pattern.

\subsection{Spurious optical polarization}
\label{subsec:Qparameters}

The presence of cross-polarization and the differences in the shape of
the polarized beams patterns from a given feed with different
orientations around its symmetry axis will result in an observation of
a polarized signal not present in the sky, but rather generated by the
instrument asymmetries. This is what we shall call ``spurious optical
polarization''. We quantify this effect in the modeled telescope as
{\it the response of the system to a diffuse unpolarized sky signal},
in terms of normalized Stokes parameters.
Following standard conventions (\cite{Kraus82}), PLANCK will measure Stokes parameters
I, Q and U by combining the output of four linearly polarized
detectors; a first pair having principal planes of polarization
at right angles to each other and a second pair with principal
planes of polarization at right angles as well, but rotated by
$45^\circ$ with respect to the first pair of detectors.
For this purpose we shall use the simulations results for the $30$
and $100$ GHz feeds at different positions in the focal plane unit
and orientations $\phi_{FP} = 0^{\circ}, 45^{\circ}, 90^{\circ},
135^{\circ}$.

The power available to the detectors is proportional
to the convolution of the beam with the sky signal, i.e.
\begin{equation}
P \propto \int_{4\pi}\!d\Omega \:(\vec{S}_B \cdot \vec{S}_S)
\label{eq:conv}
\end{equation}
where $\vec{S}_B=(I_B, Q_B, U_B, V_B)$ is the Stokes vector of the
beam on the sky and $\vec{S}_S$ is that of the sky signal. If the sky
signal is uniform, then
\begin{equation}
P \propto \int_{4\pi}\!d\Omega \:(E_{co}^2 + E_{cr}^2)
\label{eq:int}
\end{equation}
We shall use normalized Stokes parameters, $q_B = Q_B/I_B$, and $u_B =
U_B/I_B$, to characterize the {\it spurious optical polarization},
\begin{equation}
q_B =
\frac{P(\psi_1)-P(\psi_1+\frac{\pi}{2})}{P(\psi_1)+P(\psi_1+\frac{\pi}{2})}
\qquad ; \qquad
u_B = \frac{P(\psi_2)-P(\psi_2+\frac{\pi}{2})}{P(\psi_1)+P(\psi_1+\frac{\pi}{2})} \nonumber
\label{eq:Qparam}
\end{equation}
where
$P(\psi_1)+P(\psi_1+\frac{\pi}{2}) \equiv P(\psi_2)+P(\psi_2+\frac{\pi}{2})$
and $P(\psi)$ is the total power integrated over the beam
(within a co-latitude range $[-\theta_0 ; +\theta_0]$).
$\psi_2-\psi_1=45^{\circ}$ is the relative phase shift
in the polarization planes.
The calculated estimates of those parameters for the modeled feeds
are shown in Table \ref{tab:parameters} for different limits of
integration $\theta_0$. It can be seen that $q_B$ and $u_B$ depend
on the chosen integration limit, if it less than a few times the FWHM.

\begin{table*}[t!]
\small
 \centerline{\begin{tabular}{|c c c c c c c c c c|}
 \hline\hline
   {\bf 30 GHz}
   &\multicolumn{3}{c|}{Position 1}
   &\multicolumn{3}{c|}{Position 4}
   &\multicolumn{3}{c|}{\bf{Position 27}}\\ \hline
   $\theta_0$
   &$0.5^{\circ}$&$3.0^{\circ}$&$20.0^{\circ}$
   &$0.5^{\circ}$&$3.0^{\circ}$&$20.0^{\circ}$
   &$\bf{0.5^{\circ}}$&$\bf{3.0^{\circ}}$&$\bf{20.0^{\circ}}$
   \\\hline
   $q_B$ (\%)
   & -0.0391 & -0.275 & -0.275
   &  1.79   &  3.17  &  3.10
   & \bf{-0.0057 }& \bf{-0.131} & \bf{-0.131} \\\hline
   $u_B$ (\%)
   &  0.00365 &  0.00117 & 0.00110
   &  5.2     &  5.1     & 5.0
   & \bf{-0.427}   & \bf{-0.204}   & \bf{-0.205}\\\hline
\multicolumn{10}{c}{}\\
\hline\hline
   {\bf 100 GHz}
   &\multicolumn{3}{c|}{Position 1}
   &\multicolumn{3}{c|}{\bf{Position 4}}
   &\multicolumn{3}{c|}{Position 27}\\ \hline
   $\theta_0$
   &$0.2^{\circ}$&$3.0^{\circ}$&$20.0^{\circ}$
   &$\bf{0.2^{\circ}}$&$\bf{3.0^{\circ}}$&$\bf{20.0^{\circ}}$
   &$0.2^{\circ}$&$3.0^{\circ}$&$20.0^{\circ}$
   \\\hline
   $q_B$ (\%)
   &  0.164 &  0.093  &  0.093
   & \bf{-0.138} & \bf{-0.0398} & \bf{-0.0422}
   &  1.42  &  2.05   &  2.05 \\\hline
   $u_B$ (\%)
   & -0.00418 &  0.0052 &  0.0052
   & \bf{-0.51}    & \bf{-0.086}  & \bf{-0.086}
   &  2.22    &  3.44   &  3.43 \\\hline
 \end{tabular}}
\caption[] {\small{Normalized Stokes parameters $q_B$ and $u_B$ for
   different limits of integration $\theta_0$; at FWHM, $3^{\circ}$
   and $20^{\circ}$. These parameters represent the optical spurious
   polarization introduced by PLANCK telescope optics for the studied
   focal plane configuration (in bold) and other study cases (see text).}}
\label{tab:parameters}
\end{table*}

\begin{figure}[p!]   
 \centering
 \mbox{\raisebox{1.1cm}{(a)}
       \psfig{file=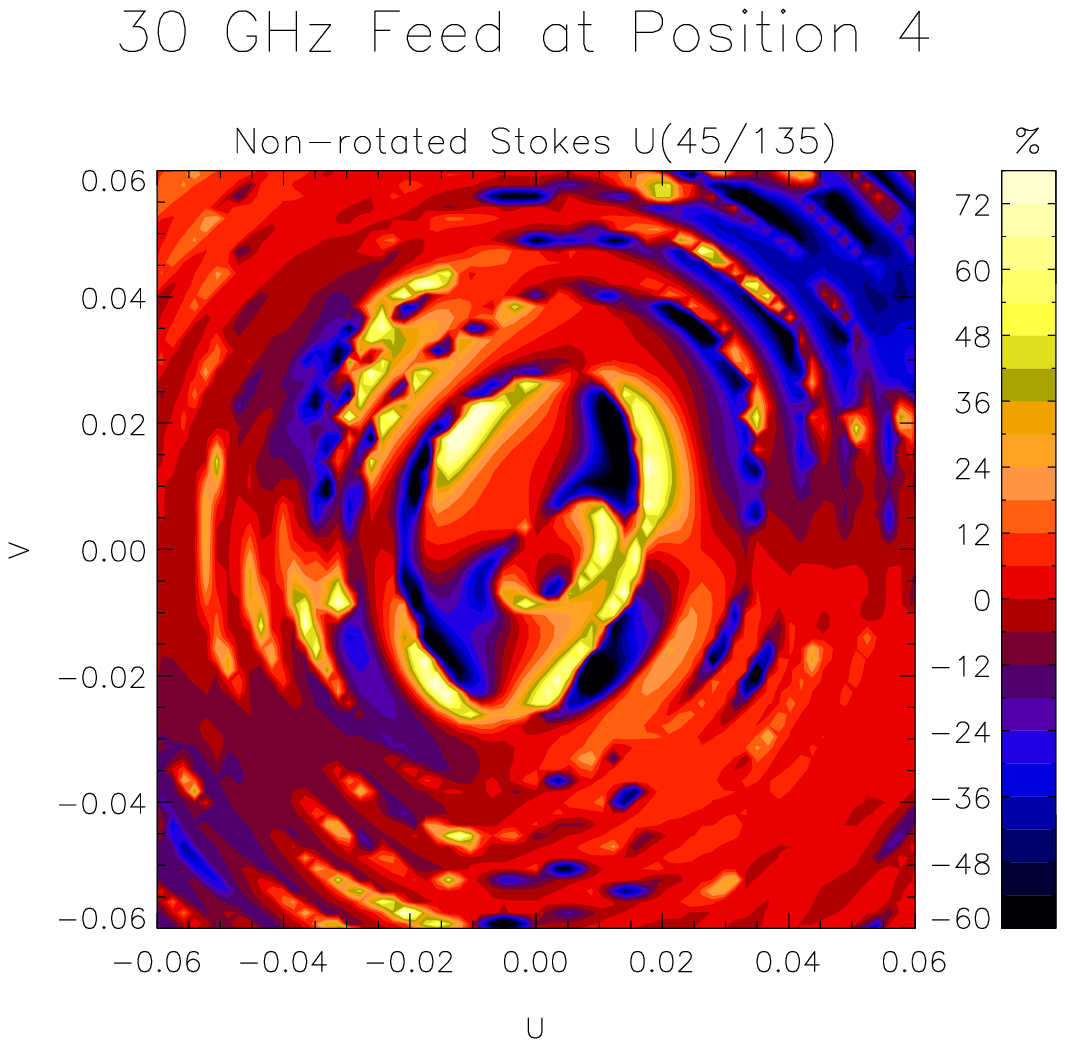,width=0.4\textwidth,clip=,silent=}
       \qquad\qquad
       \raisebox{1.1cm}{(b)}
       \psfig{file=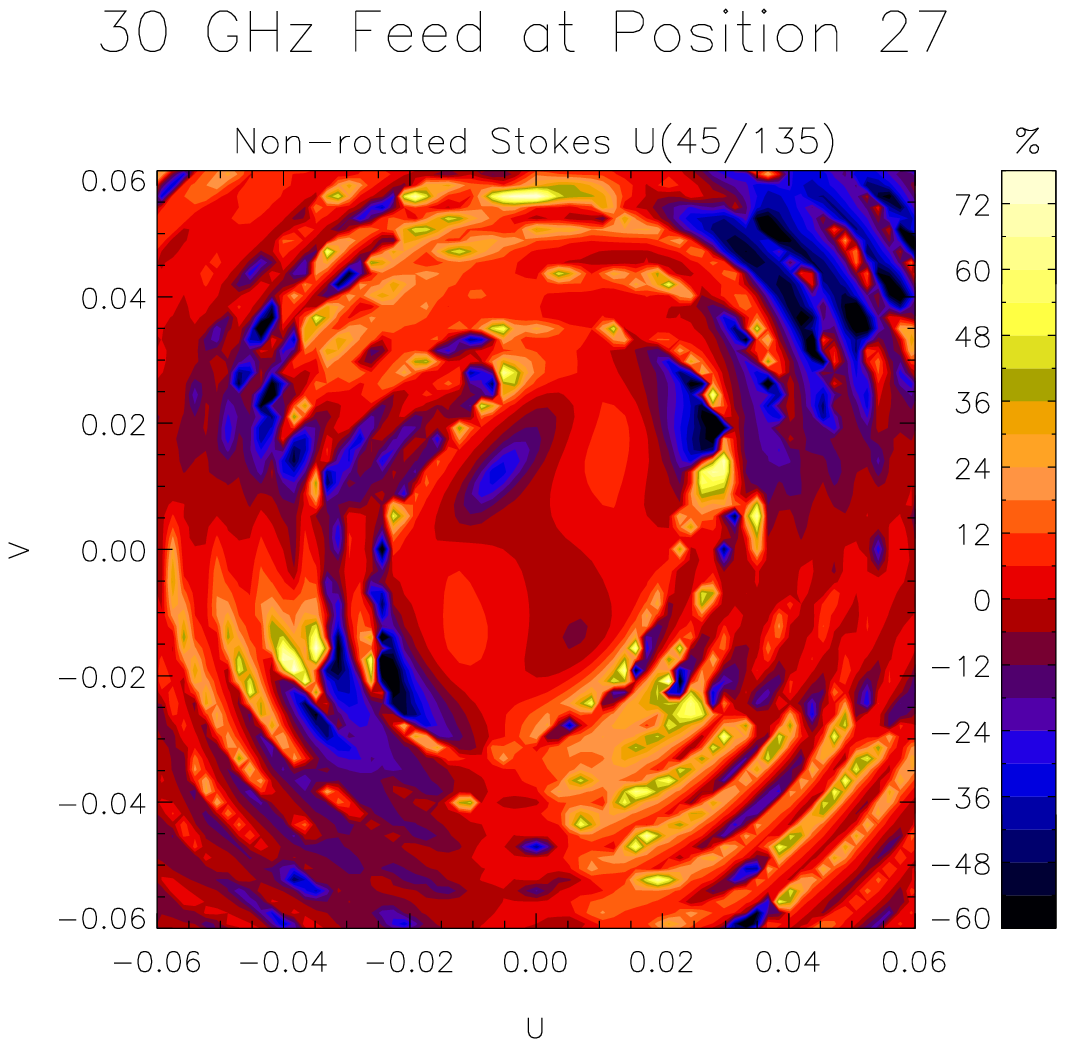,width=0.4\textwidth,clip=,silent=}}\\
 \bigskip
 \mbox{\raisebox{1.1cm}{(c)}
       \psfig{file=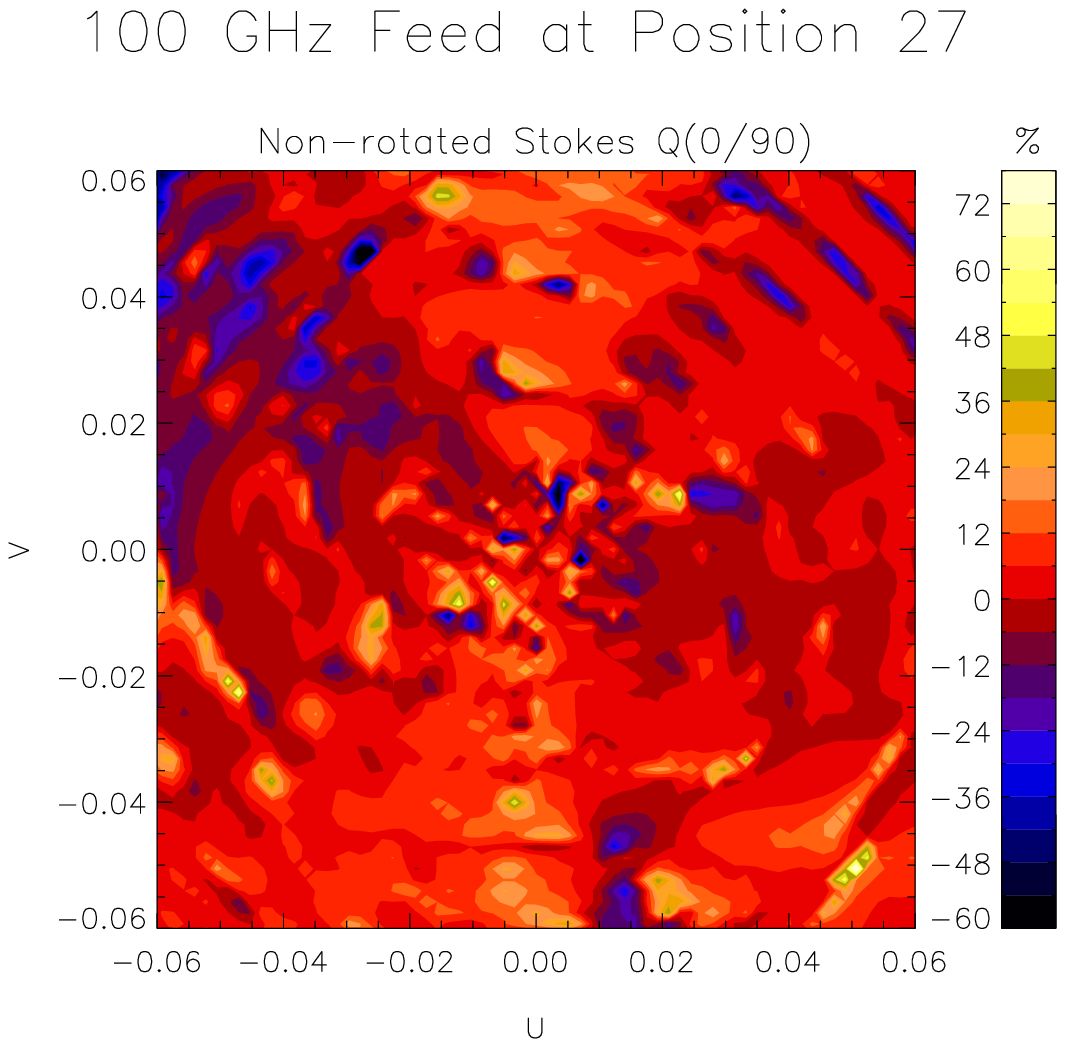,width=0.4\textwidth,clip=,silent=}
       \qquad\qquad
       \raisebox{1.1cm}{(d)}
       \psfig{file=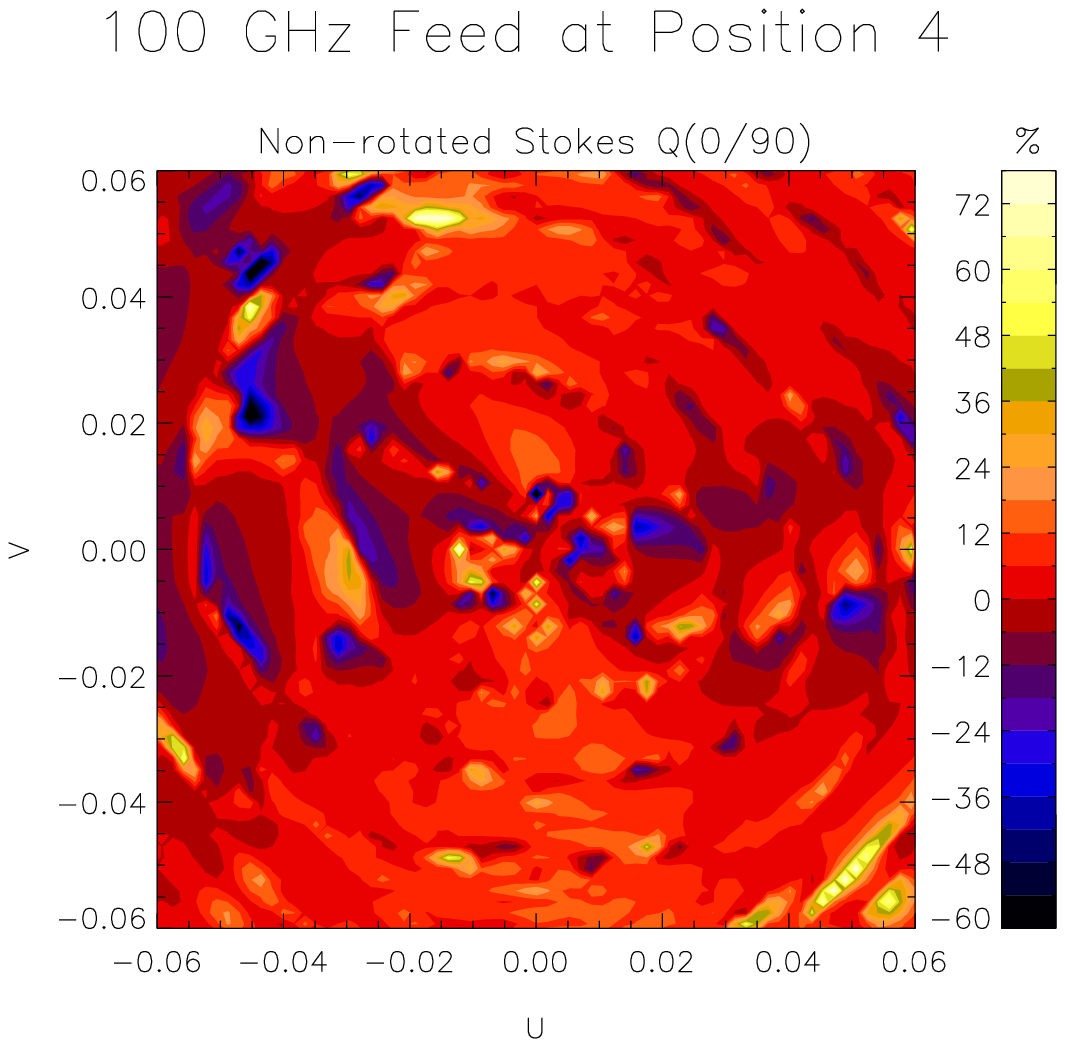,width=0.4\textwidth,clip=,silent=}}
\caption[]{Contour plot of the distribution of the normalized Stokes
   parameters across the main beam (with~$|\theta|<3^{\circ}$). The plot
   shows one of the parameters ($u_B$ or $q_B$) for the two worst study
   cases in Table~\ref{tab:parameters}: (a) the 30~GHz feed on position 4
   and (c) the 100~GHz feed on position 27. The same parameters are also
   shown for the feeds in their real positions in the studied PLANCK focal
   plane layout: (b) the 30~GHz feed on position 27 and (d) the 100~GHz
   feed on position 4.}
\label{fig:paramcontours}
\end{figure}

Table~\ref{tab:parameters} also shows that the spurious polarization
depends strongly on location and frequency and that a high level of
up to $5\%$ polarized contamination of the diffuse unpolarized intensity
is found for feeds at $30$ and $100$~GHz.  The observed large variation
of this level across the focal plane seems, at first glance, difficult
to explain. In order to illustrate the issue, we show in
Figure~\ref{fig:paramcontours} the dependence of $q_B$ and
$u_B$ across the main beam for the two worst cases in
Table~\ref{tab:parameters}; the 30~GHz feed on position~4
(Fig.~\ref{fig:paramcontours}(a)) and the 100~GHz feed on position~27
(Fig.~\ref{fig:paramcontours}(c)) and for the cases where the two
feeds are in their actual positions in the studied PLANCK focal plane
layout; the 30~GHz feed on position~27 (Fig.~\ref{fig:paramcontours}(b))
and the 100~GHz feed on position~4 (Fig.~\ref{fig:paramcontours}(d)).

It can be seen that $q_B$ and $u_B$ exhibit large fluctuations across
the main beam, which may cancel out more effectively when
integrated in angular space. The final integrated values of $q_B$
and $u_B$ will therefore depend on the fine details of the modeled
patterns, which are affected by the exact location of the feed in
the focal plane.  We can expect that, in a real system where the
detectors are sensitive to wide bandwidths, part of the fluctuations
will be averaged out. In addition, it is not clear to what extent the
fine details of the models will be reproduced by the real physical
systems. For these reasons we can consider that the levels of spurious
optical polarization here estimated are very conservative upper limits.

It is however emphasized that, given the expected small relative
amplitude of the polarized with respect to the unpolarized CMB
signal (about~$10\%$), this upper limit of optically-induced
polarization cannot be ignored. At these frequencies, there are
two large scale sources of unpolarized diffuse components in the sky:
the isotropic CMB emission (at~T~$\simeq 2.73$~K) and the CMB dipole,
a Doppler temperature anisotropy in the sky of amplitude $\sim3m$K
(see, e.g. \cite{Lineweaver96}).
As PLANCK scans the sky, spurious optical polarization will
introduce a constant bias for the former and a slowly varying
bias for the latter. The level of these biases ($\sim 30m$K and
$0.03m$K, for each 1\% of spurious polarization) is quite large
compared to the expected CMB polarized signal of $\sim 5-10\mu$K.
Note however that a constant bias does not affect the measurements
of small scale anisotropies, and a slowly varying bias is likely
to be easily removable using calibration procedures.

The main source of concern in the detection of polarized sky signals
for a PLANCK-like experiment will arise from the spurious signal that
leaks from the measured unpolarized intensity itself.
The percentage of spurious polarization is, in the very
worst cases, only factors of a few smaller than the percentage of
polarization signal. Potentially this spurious level will ultimately
introduce a fundamental limit on the measurement of CMB polarization
anisotropies at sub-degree scales, i.e. at angular scales
comparable to the main-beam size. It is however emphasized that
the spurious polarization estimated for the actual positions of the
PLANCK feeds is much lower than in the worst case scenario, and it is
certainly below the expected polarized CMB signal. Moreover, this
situation should be further improved if we take into account other
relevant effects such as the large detector bandwidth that tends to
suppress to a large extent the small-scale angular variations observed
on monochromatic radiation patterns.

\subsection{Model uncertainties}
\label{subsec:effects}

This study has not taken into account imperfections in the optics
due to the effects of dust contamination, microcracks, asymmetry
of the focal plane, or the effects of optical misalignment, since
it is today very difficult to model such effects. However,
qualitatively it can be argued that:

\begin{itemize}
\item As long as dust grains are not elongated and preferentially
  aligned, their effect is largely to scatter power from the main beam
  into the far sidelobes. In this study we are interested in the
  polarization introduced through the main beam and therefore dust
  can be safely ignored.

\item The effect of microcracks is qualitatively similar to that of
  dust with the exception that elongation and preferential
  alignment are expected. Fortunately, in the case of PLANCK, all technological
  indications are that no microcracks will be present.

\item Misalignment introduces additional aberration in the optics that
  translates into distortion of the main beam. Distortion is always
  related to depolarisation and therefore it is of potential concern.
  However, the aberrations introduced by misalignment are, by design
  and test, kept at a level which is a small fraction of the intrinsic
  optical distortions. Therefore it is reasonable in this
  initial study to assume that it will remain a second order effect.

\item An asymmetric focal plane will likely introduce differential
  effects in the response of a single horn to different planes of
  polarization. However, the effects of mutual coupling between
  PLANCK horns that are adjacent in the focal plane are kept, by
  design, at a rather low level with respect to the main beam (i.e,
  at least 30 dB below peak) and we assume here that they are not
  affecting the main results of this study.
\end{itemize}

In PLANCK, the higher frequency detector chains include also a number
of optical elements (such as filters, detector embedding structures,
cavities, etc.) which potentially introduce polarization effects.
Such chains have been measured and found to introduce very low levels
of cross-polarization. However, since they are of the same order as
the effects listed above, they should be included in future simulations.

Finally, we emphasize that this study is based on a numerical
model, which naturally is subject to some degree of uncertainty
when compared to reality. In this respect, the software used for
our simulations (GRASP) is today considered to be the benchmark
software tool for analysis of large antenna systems. Recent
studies (\cite{Murphy01}) have been carried out to investigate the
reliability of a variety of software tools, including GRASP8, in
modelling optical systems in the submillimetre wave band. In these
studies, simulated results from different software packages were
compared to each other as well as to experimental measurements.
These studies show GRASP to be the most reliable package, agreeing
with experimental measurements down to a $-40$ dB level in
predicting the amplitude pattern and showing the best agreement in
phase measurements. Moreover, the same study has shown that GRASP8
is the only available package capable of handling polarization in
the case of off-axis reflectors systems.

Therefore, in terms of model uncertainties, it is considered that
this study is as close to real as can be expected, within the
assumptions made. However, it is acknowledged that the real system
performance must ultimately be measured. A comprehensive
pre-launch test campaign is planned for PLANCK; these measurements
will be later combined with in-flight characterisation to obtain
the final system performance.

\section{Summary and Conclusions}
\label{sec:conclusions}

We have studied the systematic effects introduced by the PLANCK optical
system in the measurement of polarized signals.
For the purpose of this study we have carried out PO simulations of the
former PLANCK telescope configuration with the $30$ and $100$~GHz feeds at
three different feed locations (see Figure~\ref{fig:focalplane}) and several
orientations. From the analysis of these simulations, our results
(see~\S\ref{sec:results}) yield the following main conclusions:

\begin{itemize}
\item The {\it relative orientation} of polarization planes of diffuse
  sky signals is near-perfectly preserved by the optics, so that no
  systematic effects in the measurement of polarization are to be
  expected from the feed orientations in the current baseline. This
  symmetry is only observed when a proper definition of
  cross-polarization for non-ideal antennas is used
  (see \S\ref{subsec:xpd}).

\item The {\it absolute orientation} of polarization planes is
  systematically rotated by the telescope.  Such a misalignment angle
  can be as large as $5^{\circ}$ and depends only on frequency, increasing
  monotonically as the feed is moved towards more off-axis positions
  (see Table \ref{tab:projections}). However, this misalignment can
  be corrected by appropriate design of the focal plane.

\item Due to the high symmetry of the telescope, it is possible
  to use a pre-computed polarization main-beam pattern for a given
  feed to predict main beam patterns (within $2^{\circ}$ in co-latitude)
  for arbitrary orientations of that feed within $\sim$3~dB
  ($\sim$15~dB) for the the co-polar (cross-polar) pattern
  (see \S\ref{subsec:beamshape}).  This will allow systematic
  studies of the polarization measurements, involving a larger
  number of feeds, by using a minimal amount of CPU time.

\item Spurious optical polarization (i.e, polarization solely
  introduced by the telescope optics) is estimated to be less
  than $0.2\%$ for the actual locations of the 30 and 100~GHz
  feeds in the studied focal plane of PLANCK
  (see Figure~\ref{fig:focalplane}).
  Other (less optimal) studied cases show spurious optical
  polarization as large as $\sim$5\% of the unpolarized sky
  signal (see Table~\ref{tab:parameters}).
  It is noted however that these estimates rely on uncertain fine
  details in the modeled patterns and therefore should be considered
  as order of magnitude only. It is likely that real experiment
  features such as broad bands, detector responsivity and optical
  imperfections would affect the results, reducing the magnitude
  of this effect. Hence these simulations should only be used as
  upper limit guidelines.
\end{itemize}

Further studies on the impact of PLANCK optics on the detection of
polarization should consider the following topics:

\begin{itemize}
\item Use of the final optical configuration of the telescope, in particular
   the focal plane, has recently undergone some modifications.

\item Take into account second order effects (already discussed in
  \S\ref{subsec:effects}) that may be important to the detection of
  polarization in the main beam region.

\item Extending the studied region further into the sidelobes, to
include any second order effects that may have an impact on
  the detection of polarization in this regions.

\end{itemize}

Finally the results of these studies will be compared to
measurements obtained on qualification models of the PLANCK optics
during a pre-launch test campaign, planned to start in 2004.

\begin{acknowledgements}

We wish to thank F. Villa, M. Bersanelli, R. Mandolesi for
significant comments on the manuscript; the referee, J.
Delabrouille for useful comments on the manuscript; 
G. Giardino, and A. Mart\'in Polegre for substantial help with the
GRASP8 modelling. PF acknowledges a post-doctoral CMBNet
fellowship from the European Commission. GF acknowledges a
scholarship from the {\em Funda\c{c}\~ao para a Ci\^encia e
Tecnologia} (FCT).

\end{acknowledgements}

\appendix

\section{GRASP8 Simulations}
\label{sec:simul}

\subsection{Simulations setup and inputs}
\label{subsec:input}

As stated before, the telescope configuration used for the present
analysis is an early version of the current PLANCK telescope design.
It corresponds to a dual reflector optical system with an offset
aplanatic geometry. The exact telescope configuration is shown in
Table~\ref{tab:teleconfig}.

\begin{table}[t!]
\centerline{\begin{tabular}{||l|c||}
 \multicolumn{2}{l}{\bf Main Reflector}\\
 \hline\hline
  Surface Shape & Off-axis Ellipsoid\\\hline
  Ellipsoid Major Diameter (2a) & 22054.938mm\\\hline
  Inter Foci Distance (2c) & 20564.583mm\\\hline
  Rim Center Offset & 1038.85mm\\\hline
  Projected Rim Shape & Circular\\\hline
  Rim Projected Aperture& 750.00mm\\
 \hline\hline
 \multicolumn{2}{l}{\bf Sub Reflector}\\
 \hline\hline
  Surface Shape & Off-axis Ellipsoid\\\hline
  Ellipsoid Major Diameter (2a) & 1641.58mm\\\hline
  Inter Foci Distance (2c) & 761.92mm\\\hline
  Rim Center Offset & 328.15mm\\\hline
  Projected Rim Shape & Elliptical\\\hline
  Rim Major Axis Projected Aperture & 509.30mm\\\hline
  Rim Minor Axis Projected Aperture & 390.30mm\\
 \hline\hline
\end{tabular}}

\caption[] {\small{Early PLANCK Telescope configuration used for the
    present study (current telescope dimensions are slightly different).}}
\label{tab:teleconfig}
\end{table}

\begin{figure}[p!]  
 \bigskip\noindent
 \centerline{\psfig{file=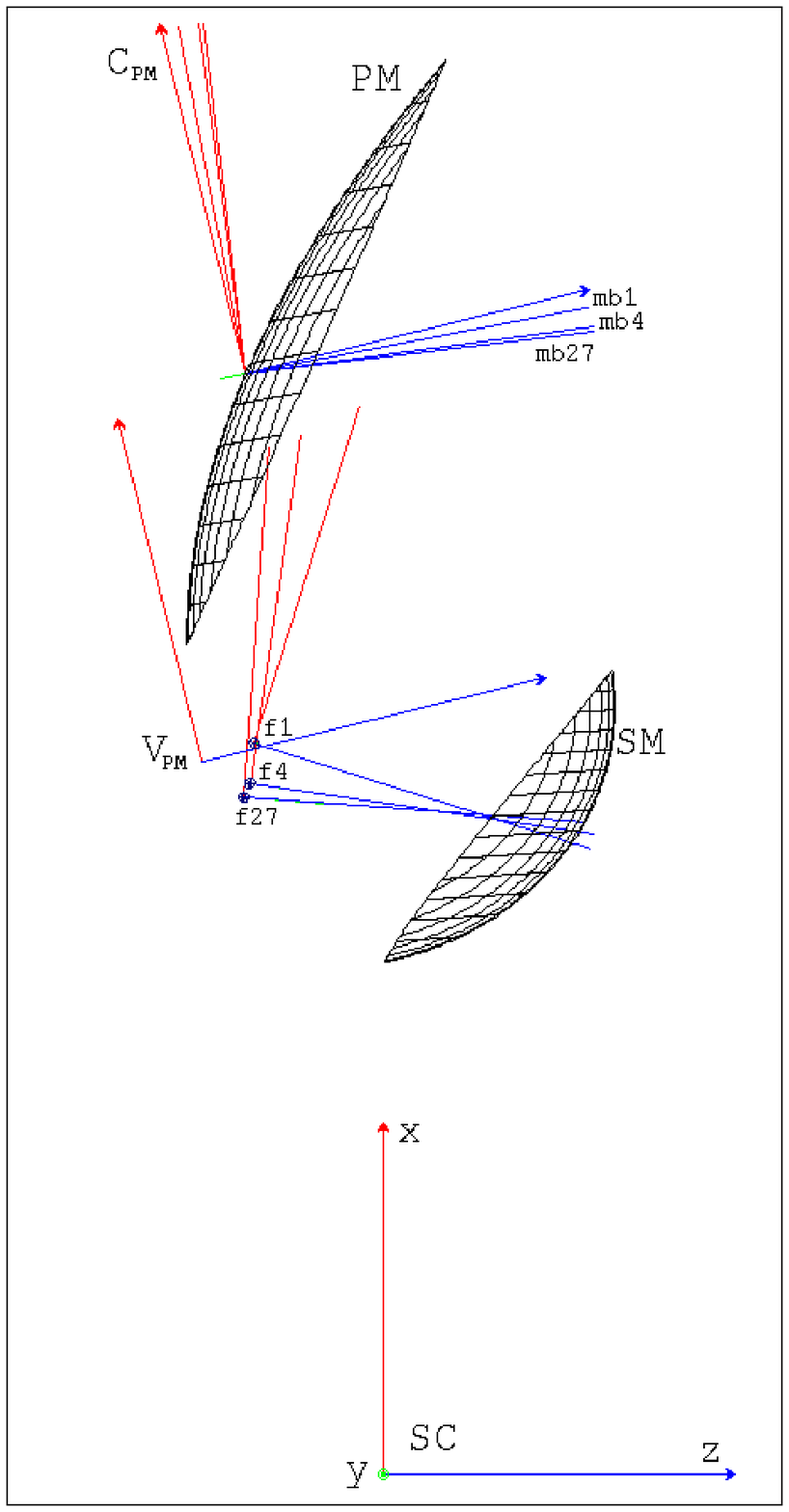,height=5.9in, clip=, silent=}}
\caption[]{Cross-section of the objects that define the Alcatel version
  of PLANCK telescope in the GRASP8 model of our study. Besides the
  primary (PM) and secondary (SM) mirrors, this plot includes the
  100~GHz feed in three different positions in the focal plane with the
  corresponding coordinate systems for the feed positions (f1, f4, f27)
  and the different orientations of the main beam in the sky
  (mb1, mb4, mb27). It also includes the fixed coordinate systems
  for the vertex ($V_{PM}$) and the center ($C_{PM}$) of the primary
  mirror and for the spacecraft (SC).}
\label{fig:side}
\end{figure}

\begin{figure}[p!]  
 \bigskip\noindent
 \centerline{\psfig{file=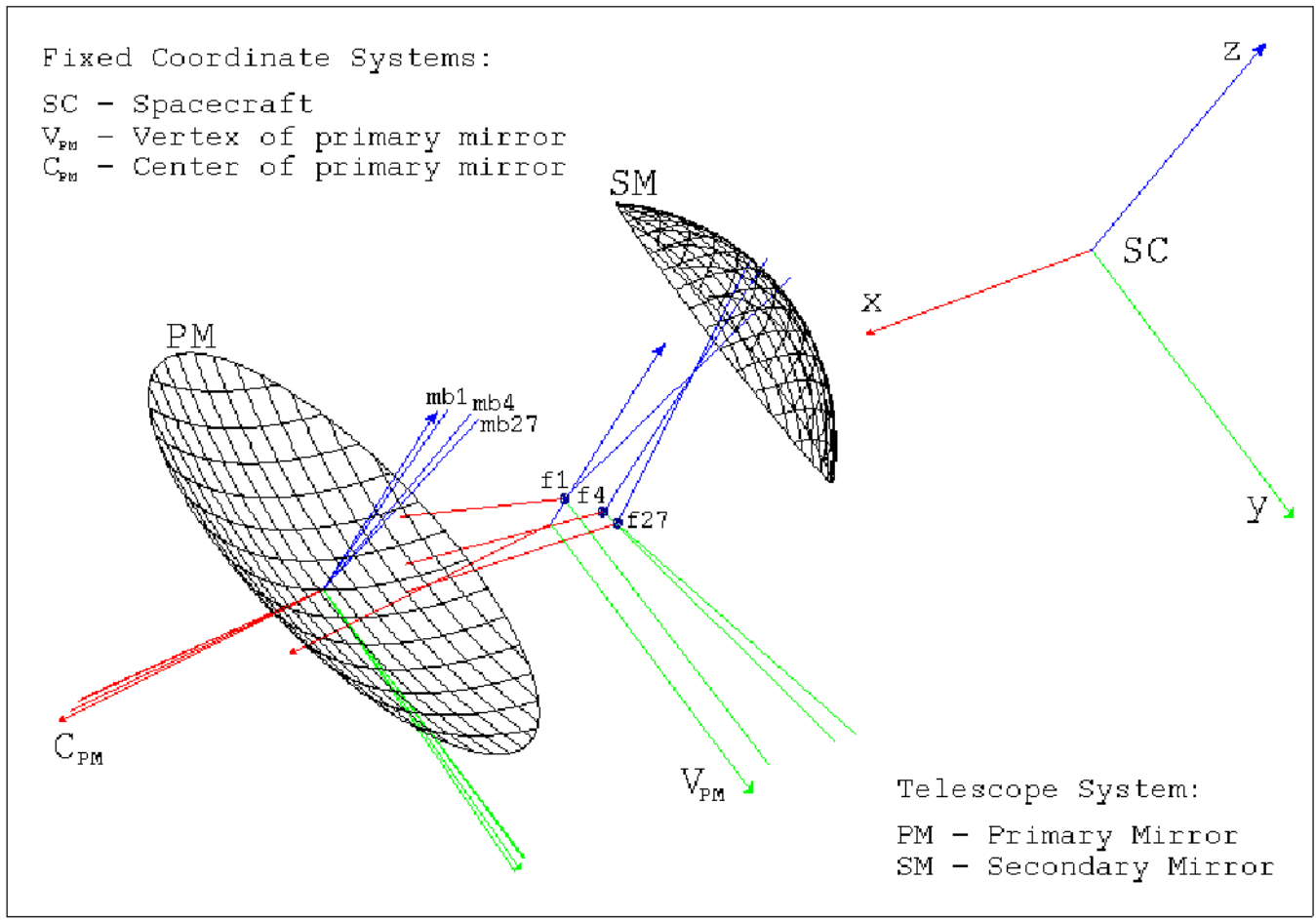,width=\textwidth,clip=,silent=}}
\caption[]{Oblique view of the objects that define the Alcatel version of
  PLANCK telescope in the GRASP8 input file of this study, as shown in
  the previous Figure~\ref{fig:side}.}
\label{fig:oblique}
\end{figure}

\begin{figure}[p!]  
 \bigskip\noindent
 \centerline{\psfig{file=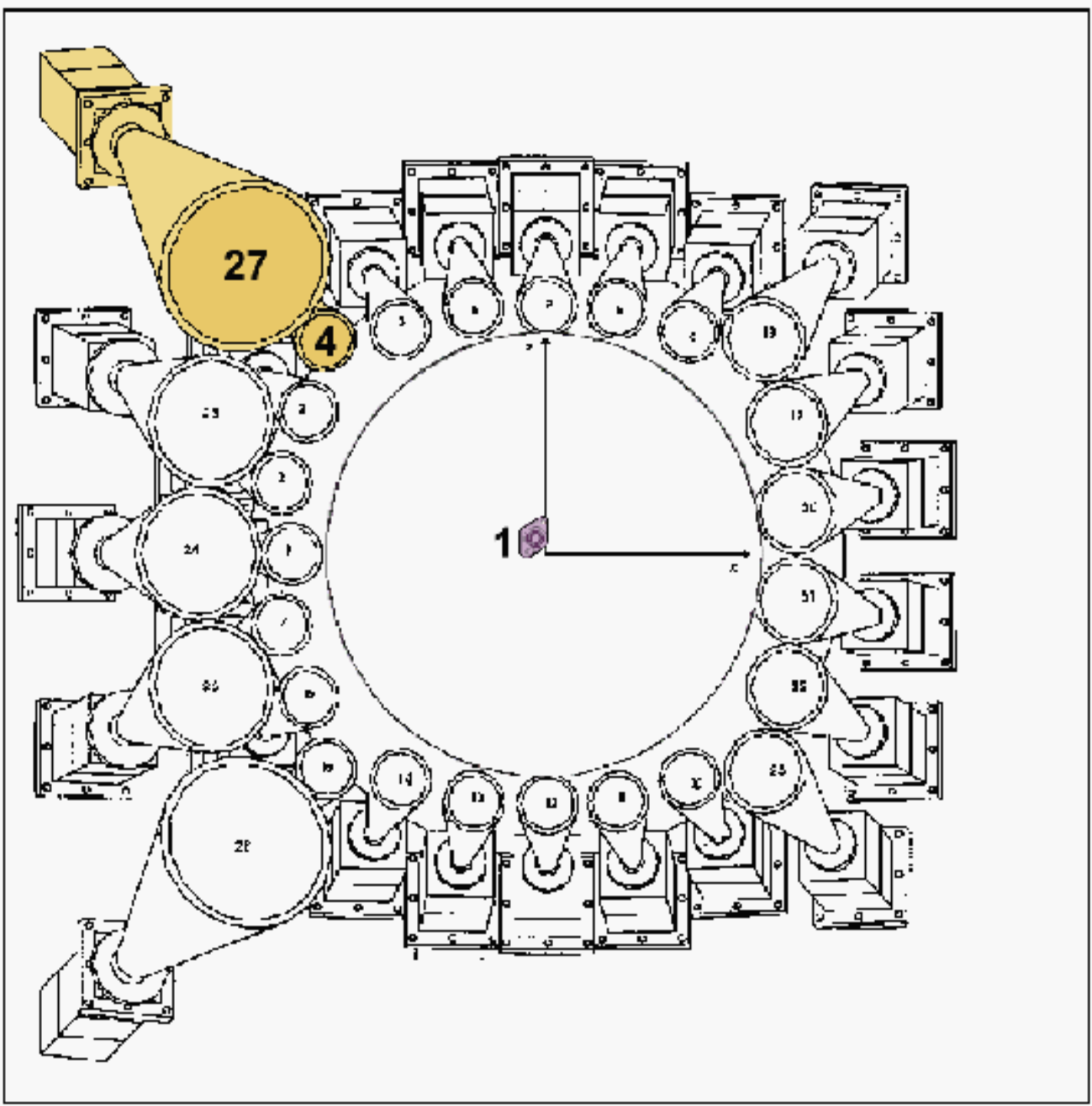,width=\textwidth,clip=,silent=}}
\caption[]{Former design of {\em PLANCK} Telescope LFI focal plane
   showing the detector positions used for this study: on the top
   left-hand corner, positions 27 and 4 with the 30 and 100~GHz
   detectors, respectively and position 1, closer to the center
   of the focal plane unit, with the HFI 857~GHz detector.}
\label{fig:focalplane}
\end{figure}

Figures~\ref{fig:side} and \ref{fig:oblique} show two different views
of the PLANCK Telescope system, with the 100~GHz feed located in three
different positions in the focal plane (positions~1, 4 and 27,
also shown with more detail in Figure~\ref{fig:focalplane}).

Reference frames f1, f4 and f27 are used to describe the three
different positions of the feed in the focal plane while reference
frames~mb1, mb4 and mb27 are used to describe the sky beam patterns,
the latter having their z-axis (in blue) always pointing to the
center of the main beam in the sky, set to be the maximum of the
co-polar pattern.
Coordinate systems f1, f4 and f27 are defined on the basis of the
coordinate system located on the vertex of the ellipse that subtends
the primary mirror ($V_{PM}$) while mb1, mb4 and mb27, are defined on
the basis of the coordinate system located on the center of the
primary mirror ($C_{PM}$) which has the same orientation as $V_{PM}$
(see Figures~\ref{fig:side} and \ref{fig:oblique}).
Furthermore, the orientation of $V_{PM}$ is the same as the spacecraft
coordinate system rotated by $+13.75^{\circ}$ around its y-axis.

For all simulations in this study, we used the polar coordinates
$(\theta,\phi)$ that can range to cover the whole sky
(i.e.,~$-\pi\leq\theta\leq\pi$ and $0\leq\phi<\pi$)\footnote{Note that
   $\theta$ is a {\em latitude} coordinate, while $\phi$ is a
   {\em longitude} coordinate.} producing ASCII data files that have
a structure such as to hold the values of the two complex amplitudes
(real and imaginary) of the two components of the beam pattern
(co- and cross-polar), for all given sets of points $(\theta,\phi)$.
All simulations were done in order to determine the radiation pattern
in the sky area around the main beam. To sample this part of the sky,
coordinate $\theta$ was chosen to range within the limits
$-20^{\circ}\leq\theta\leq+20^{\circ}$ in a set of 401 points with
intervals of $0.1^{\circ}$ while coordinate $\phi$ ranged within
$0^{\circ}\leq\phi<180^{\circ}$ in a set of 36 {\em polar cuts}
separated by $5^{\circ}$.

In a GRASP8 simulation, the scattered field can be calculated by using
Physical Optics (PO) combined with the Physical Theory of Diffraction
(PTD) or, alternatively, using Geometrical Optics (GO) combined with
the Geometrical Theory of Diffraction (GTD).
According to \cite{Ticra97}, for focused reflector systems PO/PTD
is intended to be used in the far field around beam maxima, whereas
GTD can be used in the side-lobe region. For these reasons and since
we were only interested in the main beam sky area, our simulations
were done with PO/PTD calculations.
In what concerns the density of the PO integration grid, to determine
the number of points for which the PO currents are calculated in each
reflecting surface, we referred to a formula given by \cite{Ticra97}
where the wavelength, the reflector diameter and the maximum latitude,
$\theta_{max}$, are input parameters in the determination of the
minimum number of points for which the PO integration converges.

\begin{figure}[p!]   
 \bigskip\noindent
\begin{minipage}[t]{0.49\textwidth}
 \centerline{\psfig{file=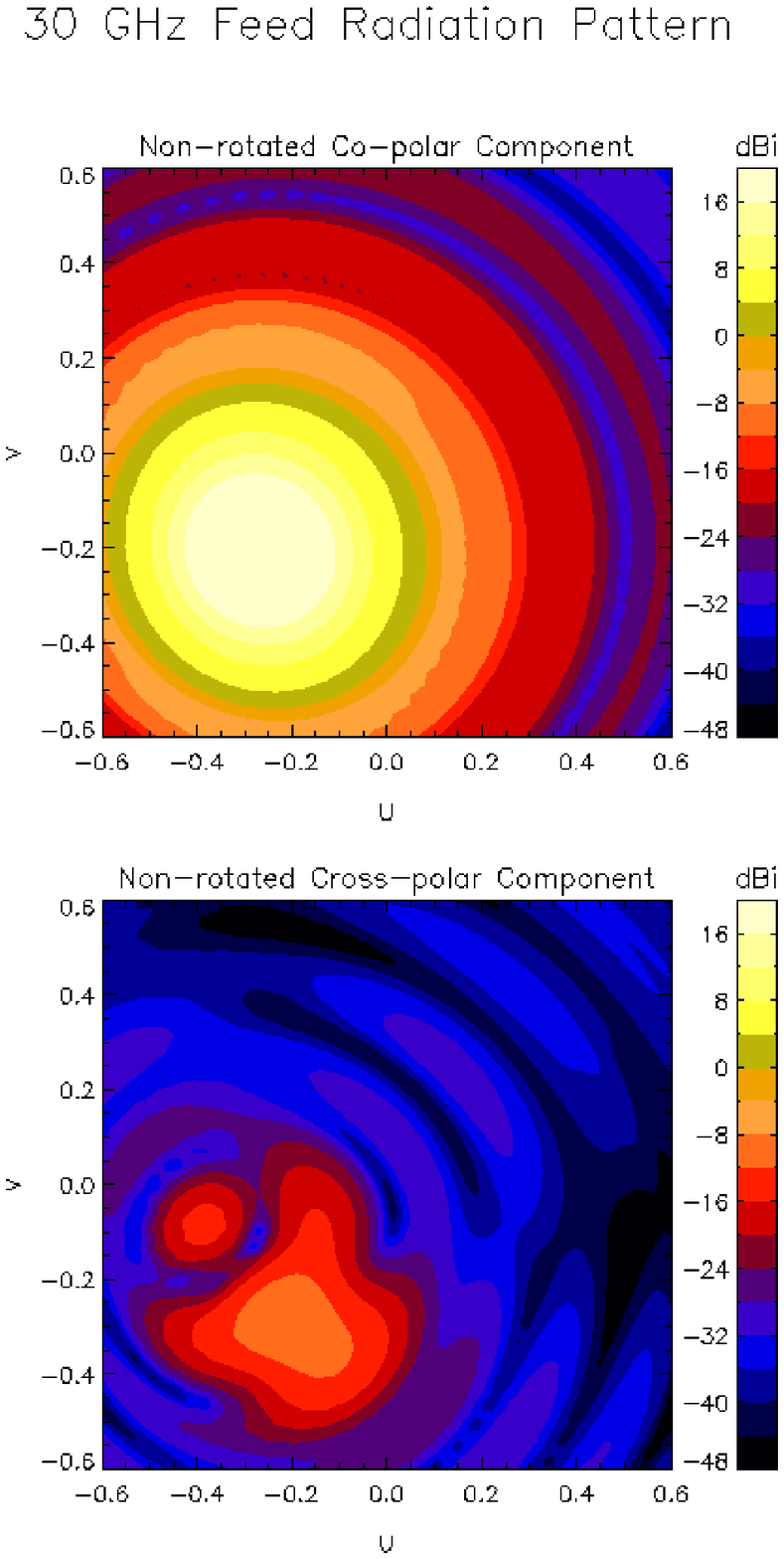,height=4.2in,clip=,silent=}}
\end{minipage}\hfill
\begin{minipage}[t]{0.49\textwidth}
 \centerline{\psfig{file=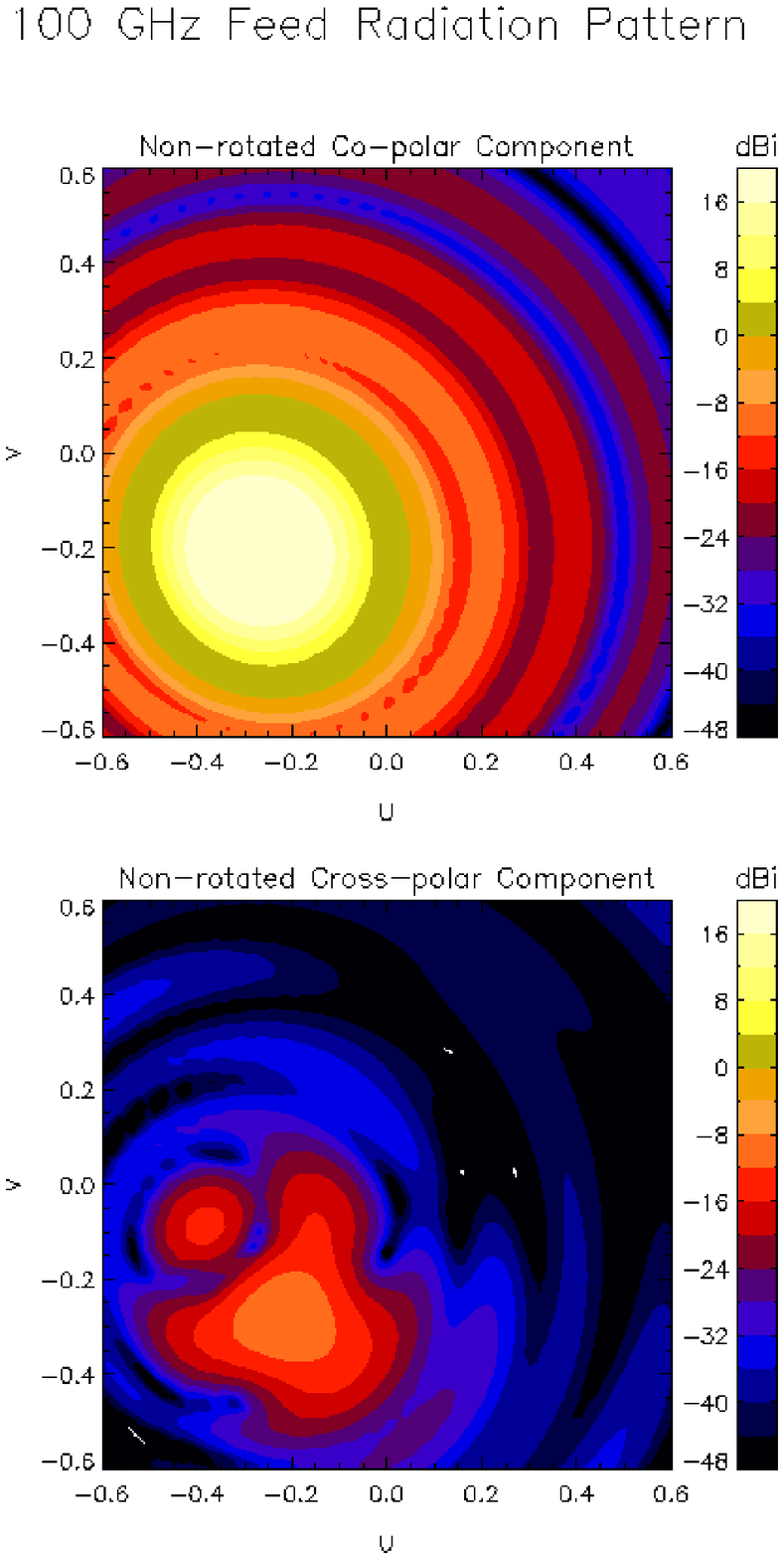,height=4.2in,clip=,silent=}}
\end{minipage}

\caption[]{Contour plots of the far field radiation pattern
   ($|\theta|<30^{\circ}$) when the feeds radiate directly to the sky,
   i.e., the input polarized signal does not go through the telescope
   system.}
\label{fig:inpattern}

 \bigskip\noindent   
\begin{minipage}[t]{0.49\textwidth}
 \centerline{\psfig{file=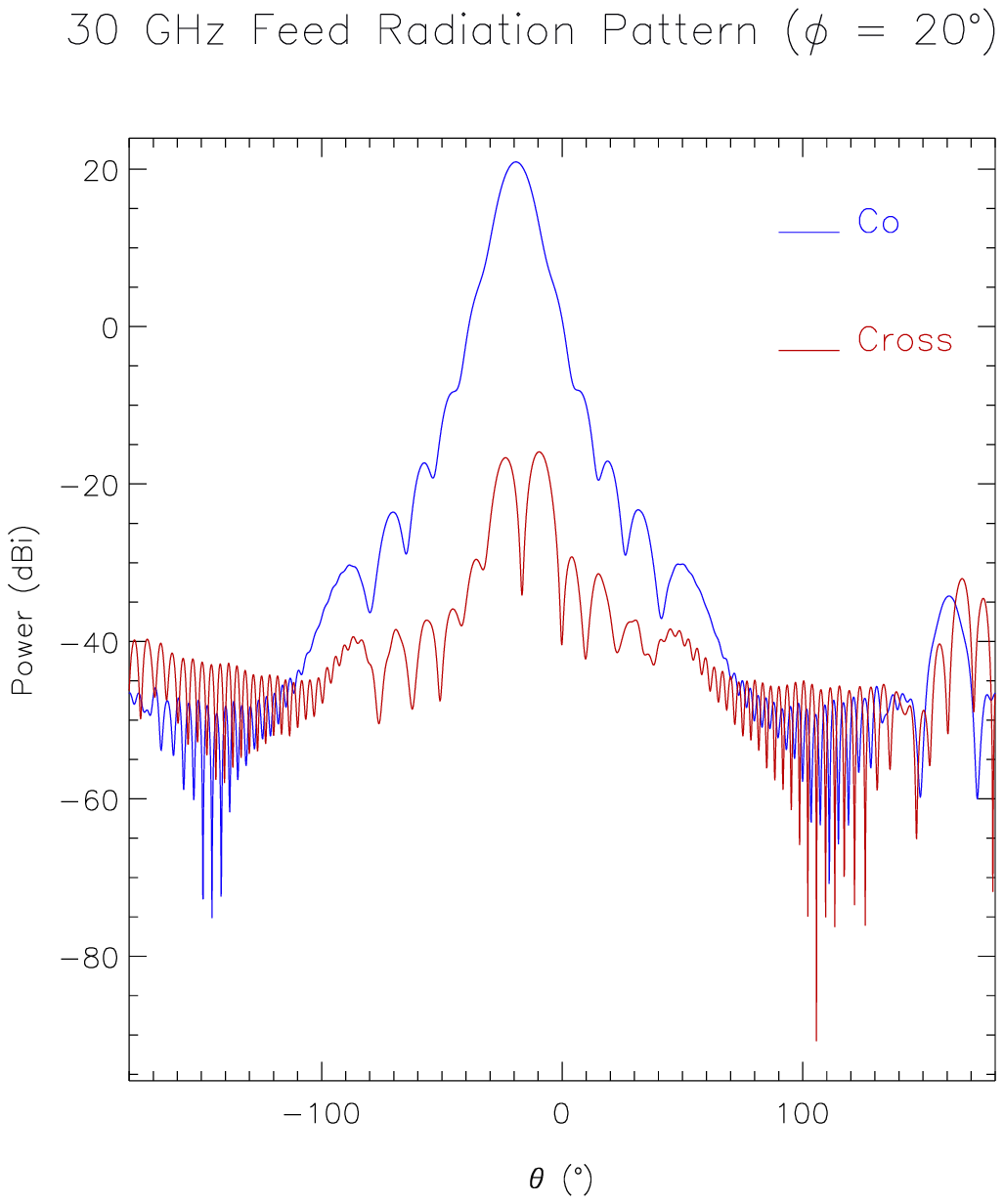,width=\textwidth, clip=, silent=}}
\end{minipage}\hfill
\begin{minipage}[t]{0.49\textwidth}
 \centerline{\psfig{file=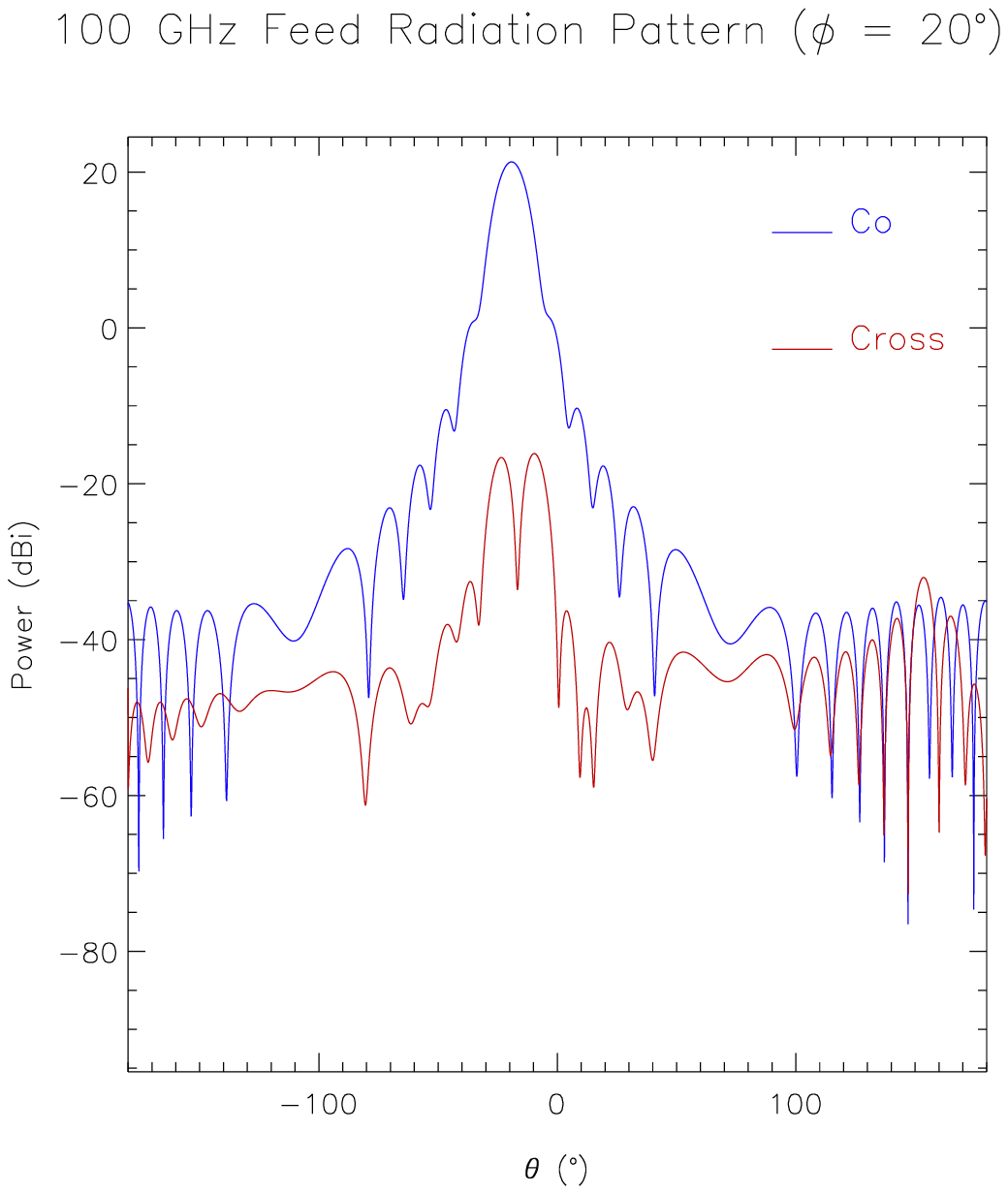,width=\textwidth, clip=, silent=}}
\end{minipage}

\caption[]{{\em Polar cuts}, at $\phi=20^{\circ}$, of the contour plots
   in Figure~\ref{fig:inpattern}.}
\label{fig:insignal}
\end{figure}

To simulate the conical corrugated horns used in the PLANCK telescope,
the input radiation signal of the feeds used in our simulations
had the form of a spherical wave expansion.
Figures~\ref{fig:insignal} and \ref{fig:inpattern} show the far field
radiation pattern when the input signal radiates directly from the
feed to the sky, without going through the telescope system;
Figure~\ref{fig:insignal} shows the plot of the {\em polar cut}
at $\phi=20^{\circ}$ while Figure~\ref{fig:inpattern} shows the
whole contour in the (U,V) plane.

Note that to project the polar coordinate system $(\theta,\phi)$
onto the $(U,V)$ plane we have the following transformation,

\begin{equation}
\begin{array}{lcr}
U=\sin\theta\cos\phi \qquad&;&\qquad -\pi\leq\theta\leq\pi\\
V=\sin\theta\sin\phi \qquad&;&\qquad 0\leq\phi<\pi
\end{array}
\end{equation}

In Figures~\ref{fig:inpattern} and \ref{fig:insignal}, note that the
beam maxima, the co-polar peaks, are not centered at $(U,V)=(0,0)$
nor at latitude $\theta=0^{\circ}$. This off-center positioning
of the beam pattern in the coordinate system $C_{PM}$ when the feeds
radiate directly to the sky is due to the off-axis positioning
of the feeds in this same reference frame.
When the radiation pattern goes through the telescope system instead
of going directly from the feed to the sky, a much smaller off-center
positioning of the beam can be observed in the coordinate system
$C_{PM}$. However, in order to have a centered beam maximum in all
plots and contours, for each of the three different feed positions
in the focal plane (1,~4~and~27), the coordinate system $C_{PM}$
was reoriented in order to always have the z-axis pointing to the
co-polarization power peak. This procedure originated the three
different coordinate systems (mb1, mb4 and mb27), already mentioned
before\footnote{These far field reference frames (mb1, mb4 and mb27)
  can be seen in  Figures~\ref{fig:side} and \ref{fig:oblique}.
  Note that they have a common origin with $C_{PM}$, at the center
  of the primary mirror, but different orientations.}.

\subsection{Simulations outputs and results}
\label{subsec:output}

\begin{figure}[p!]  
 \bigskip\noindent
\begin{minipage}[t]{0.49\textwidth}
 \centerline{\psfig{file=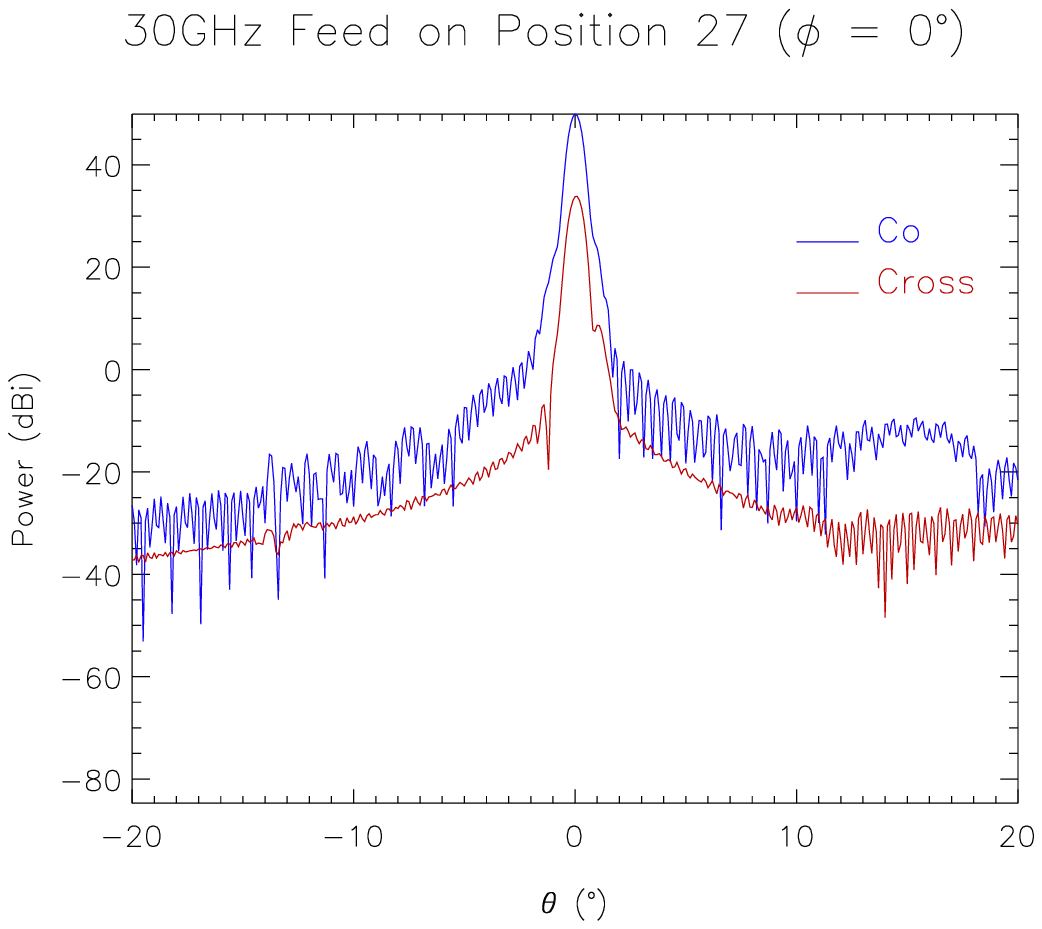,width=\textwidth, clip=, silent=}}
\end{minipage}\hfill
\begin{minipage}[t]{0.49\textwidth}
 \centerline{\psfig{file=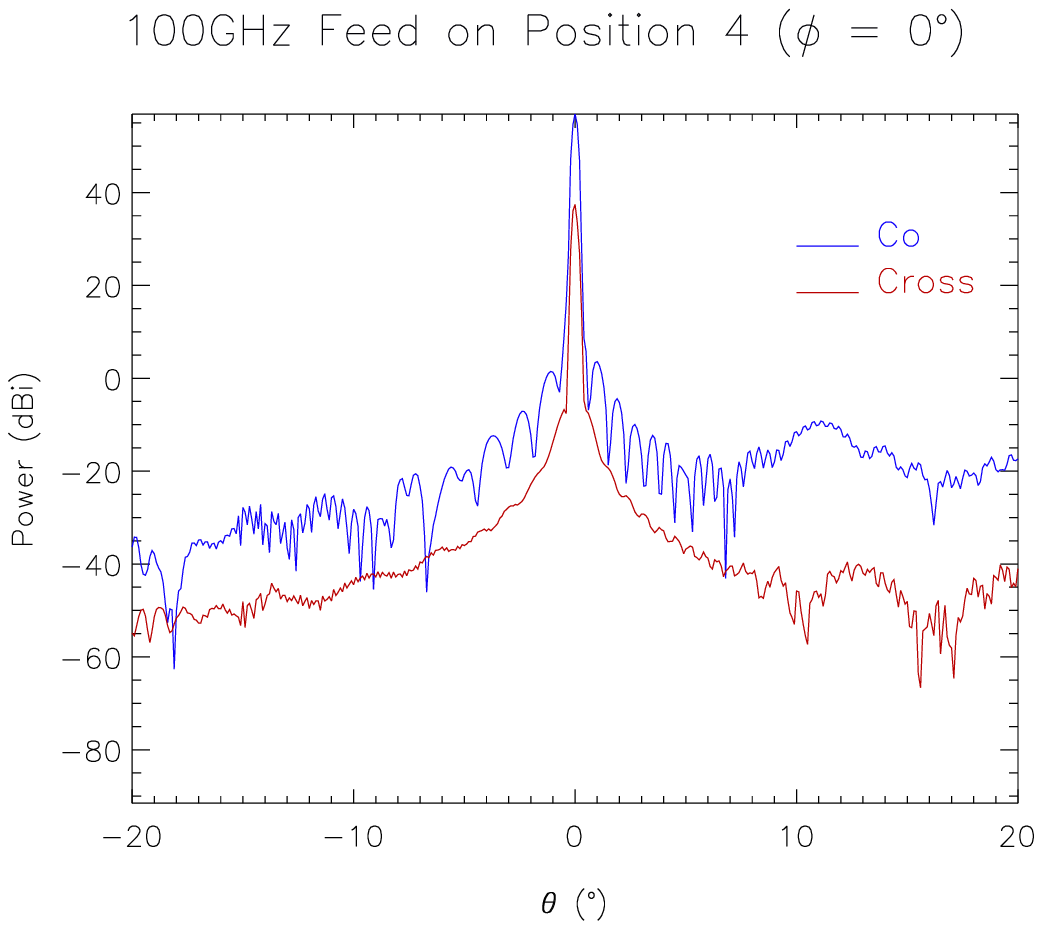,width=\textwidth, clip=, silent=}}
\end{minipage}

\caption[]{{\em Polar cuts} at $\phi=0^{\circ}$ of the 30 and 100~GHz
    sky radiation patterns when the input polarized signal goes through
    the telescope system.}
\label{fig:outsignal}
\end{figure}

\begin{figure}[p!]   
 \bigskip\noindent
  \unitlength1.0cm
\begin{minipage}[t]{8.5cm}
  \parbox[t]{0cm}{} \put(1.0,1.0){\small (a)} \put(10.5,1.0){\small
    (b)}
 \centerline{\psfig{file=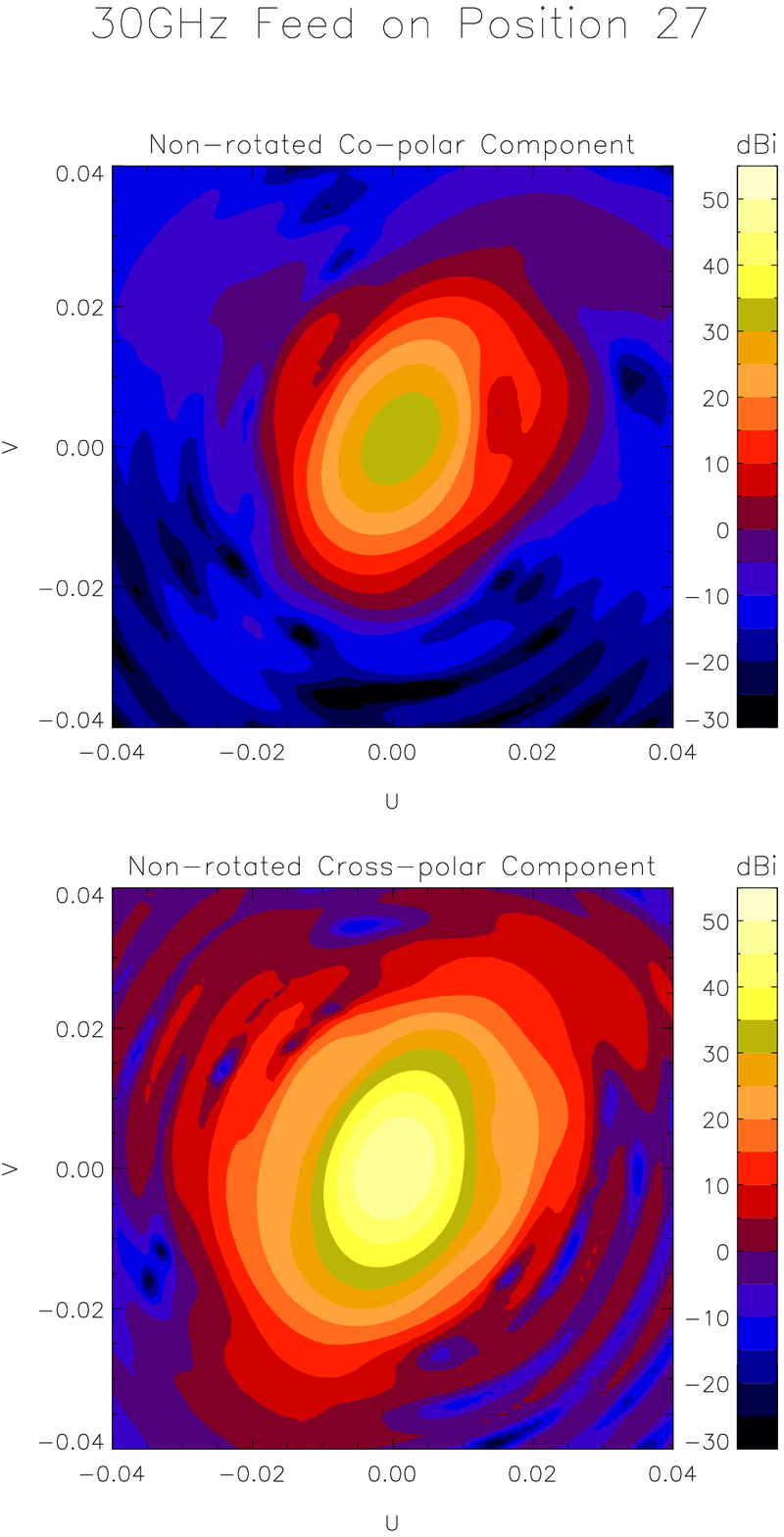,height=4.2in,clip=,silent=}}
\end{minipage}\hfill
\begin{minipage}[t]{8.5cm}
 \centerline{\psfig{file=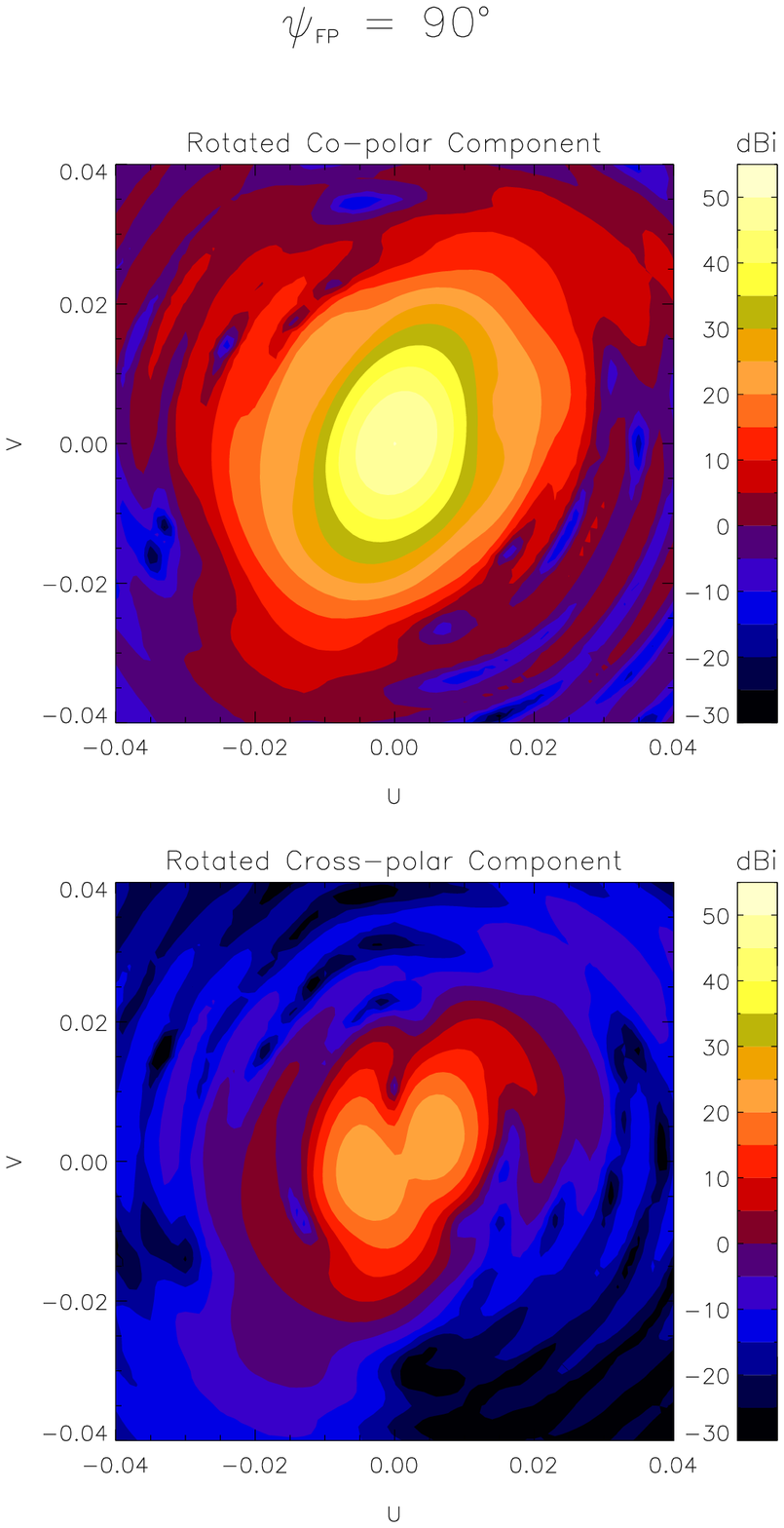,height=4.2in,clip=,silent=}}
\end{minipage}

\begin{minipage}[t]{8.5cm}
  \parbox[t]{0cm}{} \put(1.0,1.0){\small (c)} \put(10.5,1.0){\small
    (d)}
 \centerline{\psfig{file=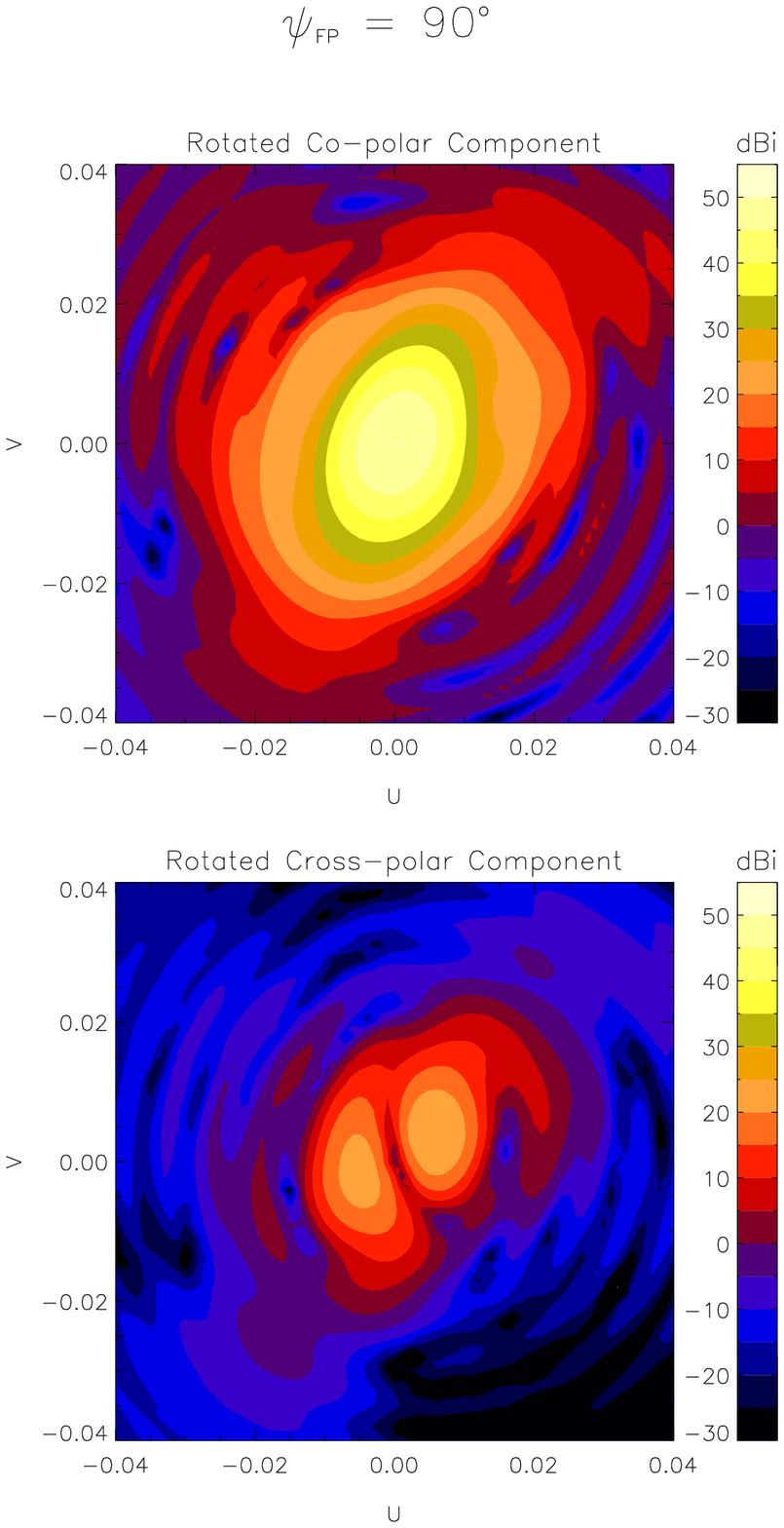,height=4.2in,clip=,silent=}}
\end{minipage}\hfill
\begin{minipage}[t]{8.5cm}
 \centerline{\psfig{file=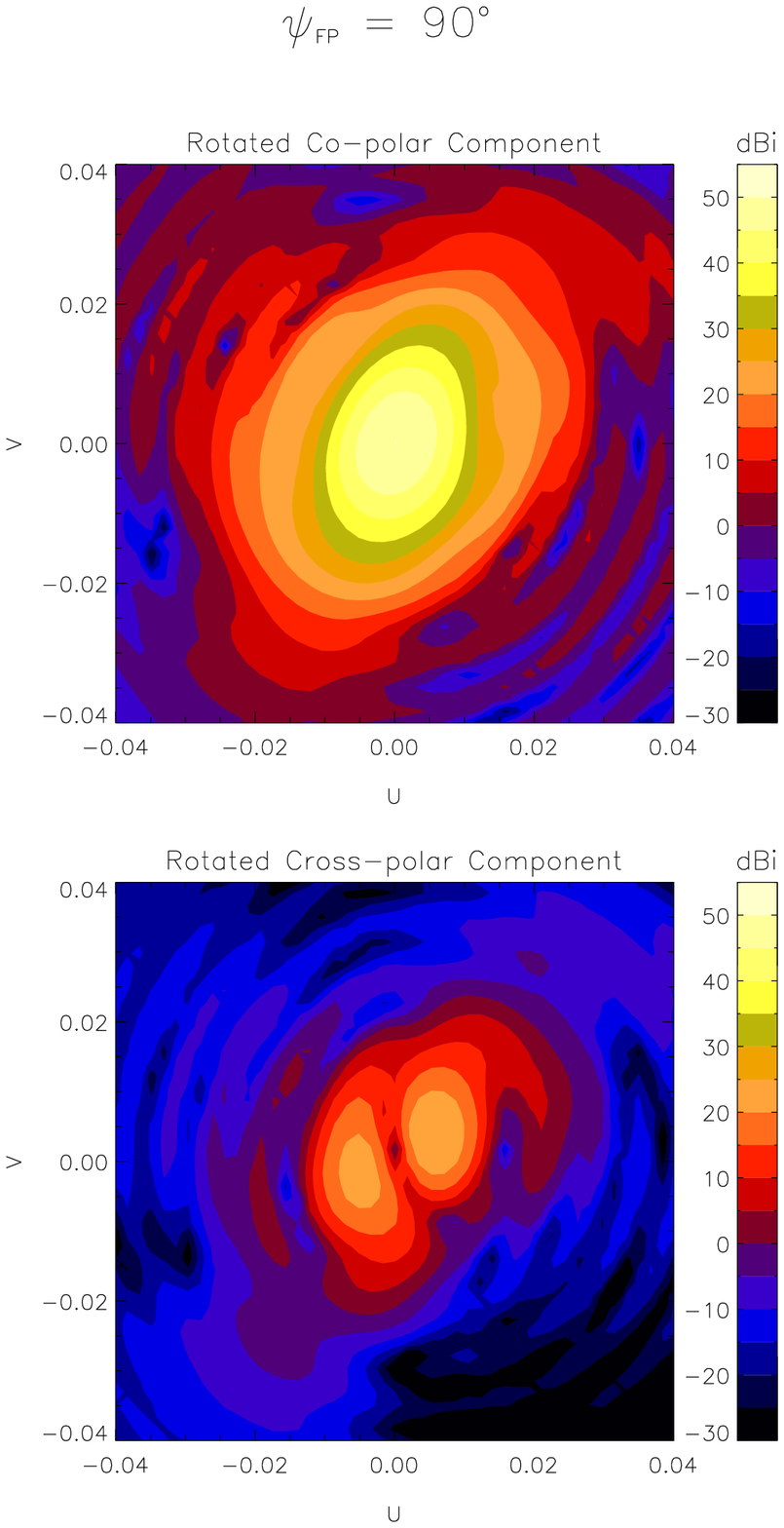,height=4.2in,clip=,silent=}}
\end{minipage}

\caption[]{Contour plots of the Co- and Cross-polar patterns around
    the main beam ($|\theta|<2^{\circ}$) for the 30 GHz feed rotated
    by $\psi_{FP}=90^{\circ}$ at position 27. The figure shows
    (a) the beam pattern in the non-rotated far field reference frame and
    the same beam pattern in the far field reference frame rotated by
    (b) $\phi_{XPD}=-79.6^{\circ}$ (see~table~\ref{tab:resultspk}),
    (c) $\phi_{XPD}=-81.5^{\circ}$ (see~table~\ref{tab:resultscp}) and
    (d) $\phi_{XPD}=-81.3^{\circ}$ (see~table~\ref{tab:resultsint}).}
\label{fig:3027psi90}
\end{figure}

\begin{figure}[p!]   
 \bigskip\noindent
  \unitlength1.0cm
\begin{minipage}[t]{8.5cm}
  \parbox[t]{0cm}{} \put(1.0,1.0){\small (a)} \put(10.5,1.0){\small
    (b)}
 \centerline{\psfig{file=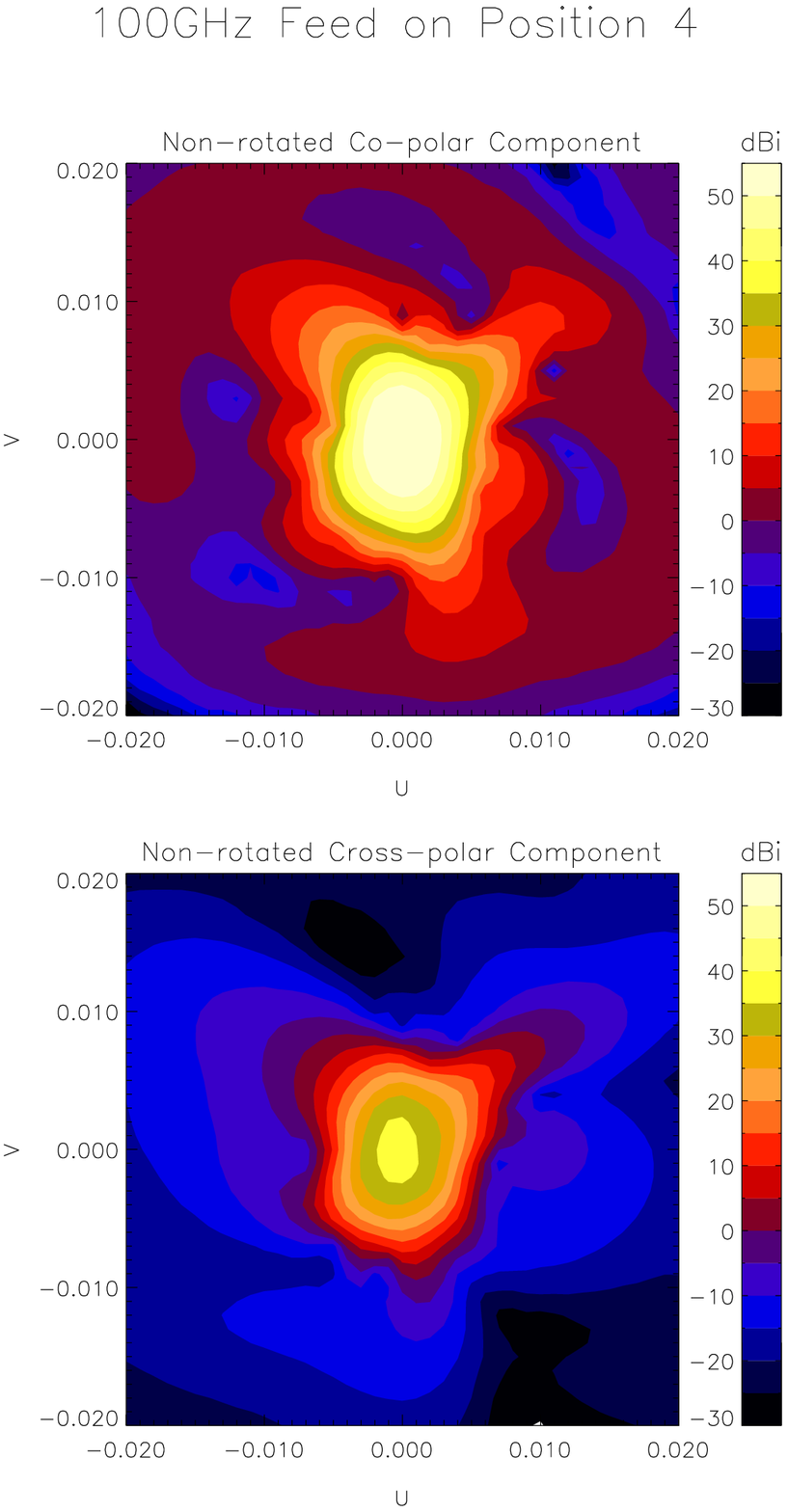,height=4.2in,clip=,silent=}}
\end{minipage}\hfill
\begin{minipage}[t]{8.5cm}
 \centerline{\psfig{file=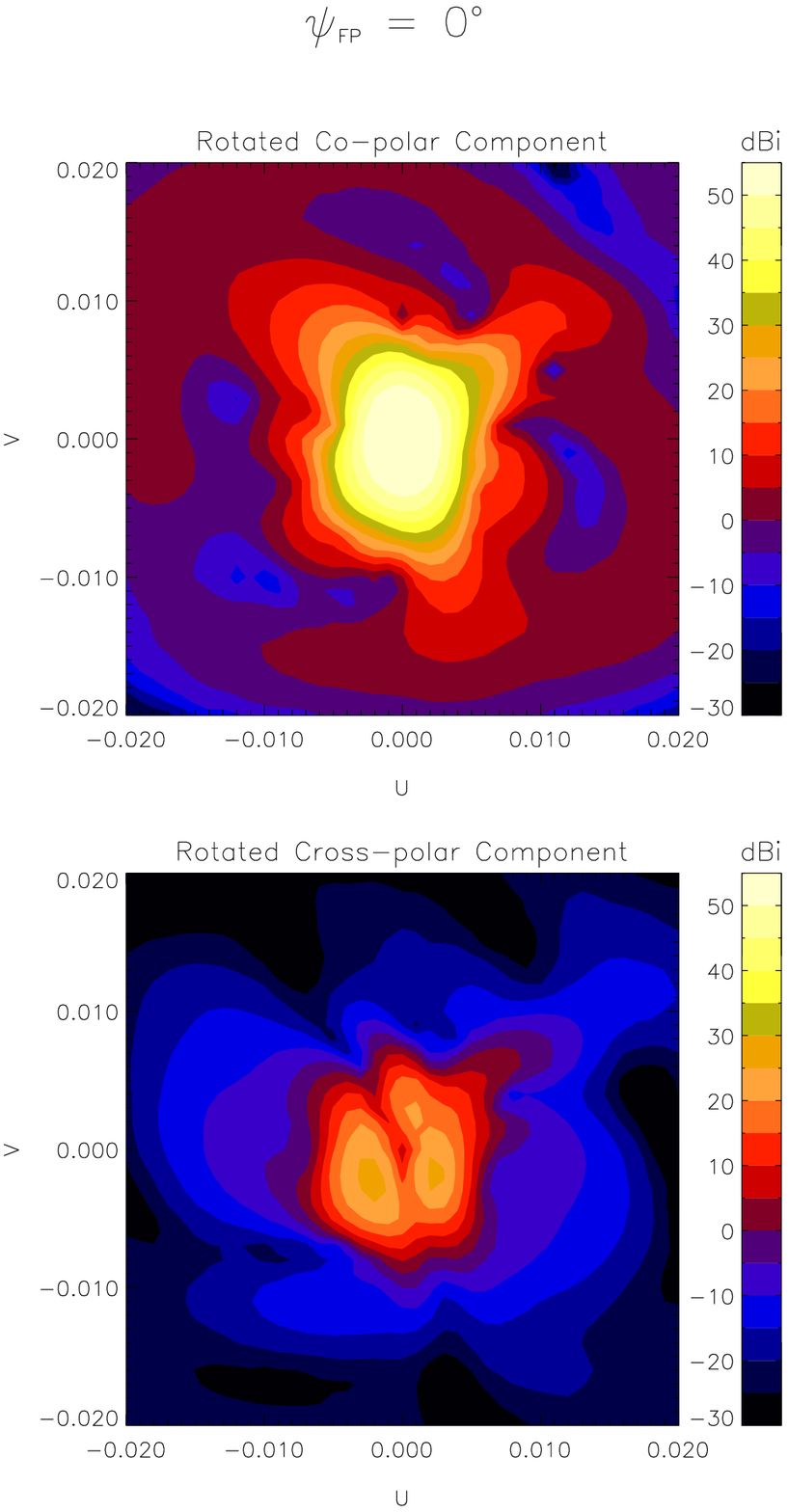,height=4.2in,clip=,silent=}}
\end{minipage}

\begin{minipage}[t]{8.5cm}
  \parbox[t]{0cm}{} \put(1.0,1.0){\small (c)} \put(10.5,1.0){\small
    (d)}
 \centerline{\psfig{file=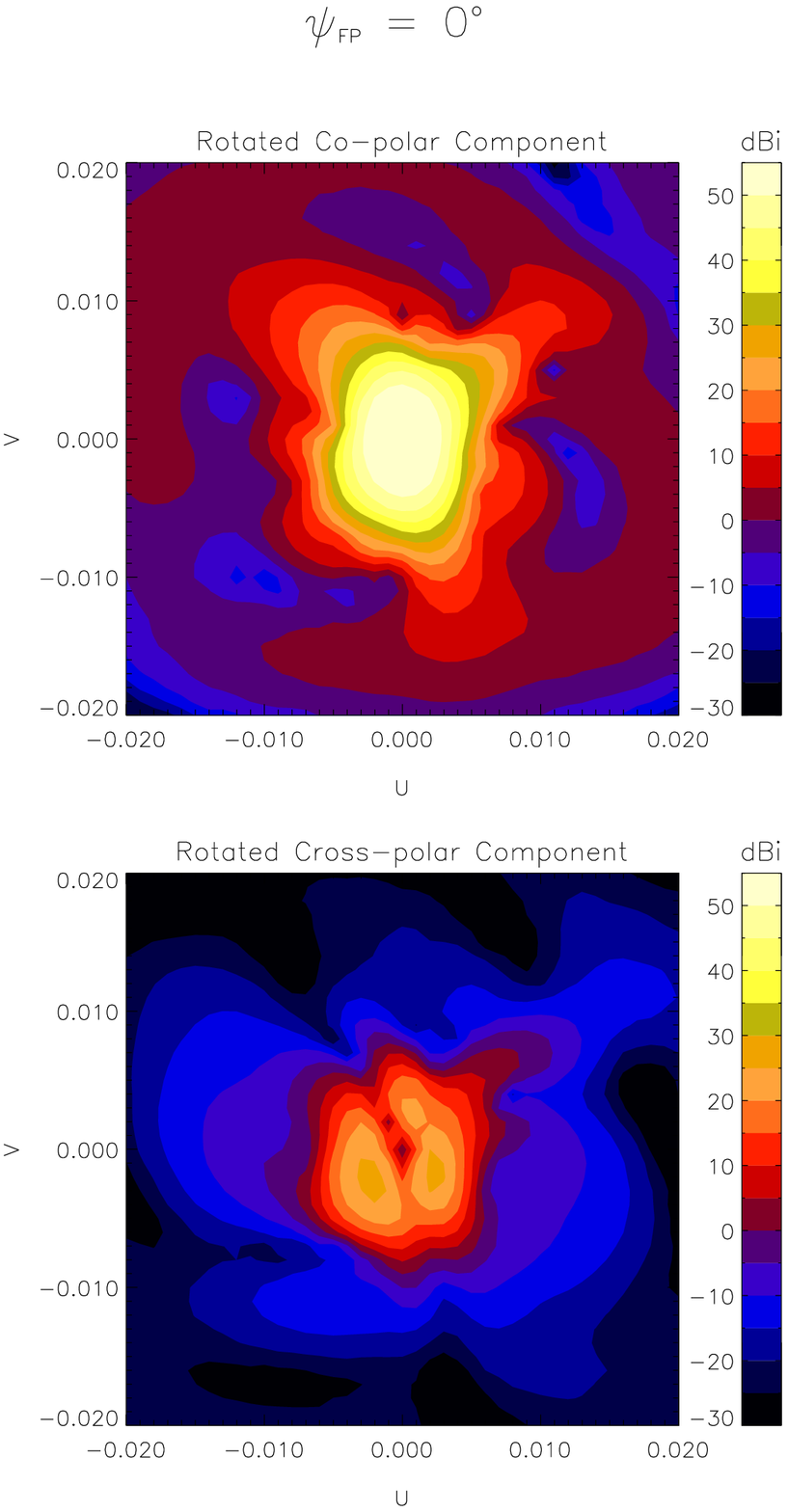,height=4.2in,clip=,silent=}}
\end{minipage}\hfill
\begin{minipage}[t]{8.5cm}
 \centerline{\psfig{file=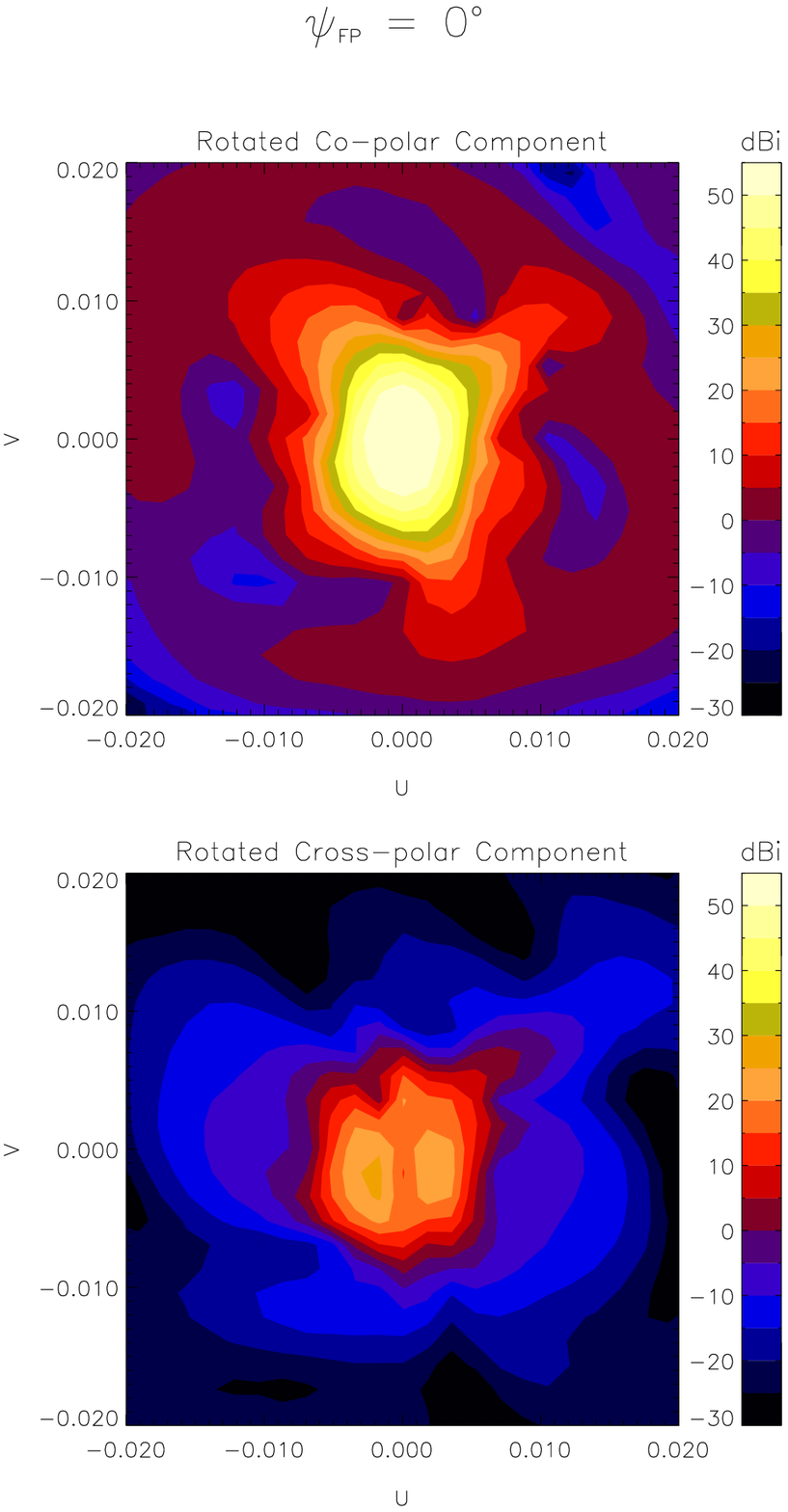,height=4.2in,clip=,silent=}}
\end{minipage}

\caption[]{Same as Figure~\ref{fig:3027psi90} (with~$|\theta|<1^{\circ}$)
    for the 100 GHz non-rotated feed ($\psi_{FP}=0^{\circ}$) at position 4.
    In this figure the far field reference frame is rotated by
    (b) $\phi_{XPD}^{nr}=5.9^{\circ}$ (see~table~\ref{tab:resultspk}),
    (c) $\phi_{XPD}^{nr}=6.1^{\circ}$ (see~table~\ref{tab:resultscp}) and
    (d) $\phi_{XPD}^{nr}=5.5^{\circ}$ (see~table~\ref{tab:resultsint}).}
\label{fig:1004psi0}
\end{figure}

Figure \ref{fig:outsignal} shows the far field radiation patterns
when the input polarized signal goes through the telescope system.
It can be seen that, as a result of the far field reference frame
reorientation mentioned in the previous section, the co-polar peaks
are always centered at $\theta=0^{\circ}$ whether the feed is on
position~27 (left plot) or on position~4 (right plot).
We show the contour plots of the far field
radiation pattern around the main beam for various simulations.
In particular, Figures~\ref{fig:3027psi90} and \ref{fig:1004psi0}, show 
radiation patterns in the non-rotated far field reference frame
and in the reference frame rotated by the $\phi_{XPD}$ angle. This
angle is determined by each one of the three different methods described in
\S\ref{subsec:frames}. Plots are shown side by side for an easier comparison.

\begin{figure}[p!]   
 \bigskip\noindent
  \unitlength1.0cm
\begin{minipage}[t]{8.5cm}
  \parbox[t]{0cm}{} \put(1.0,1.0){\small (a)} \put(10.5,1.0){\small
    (b)}
 \centerline{\psfig{file=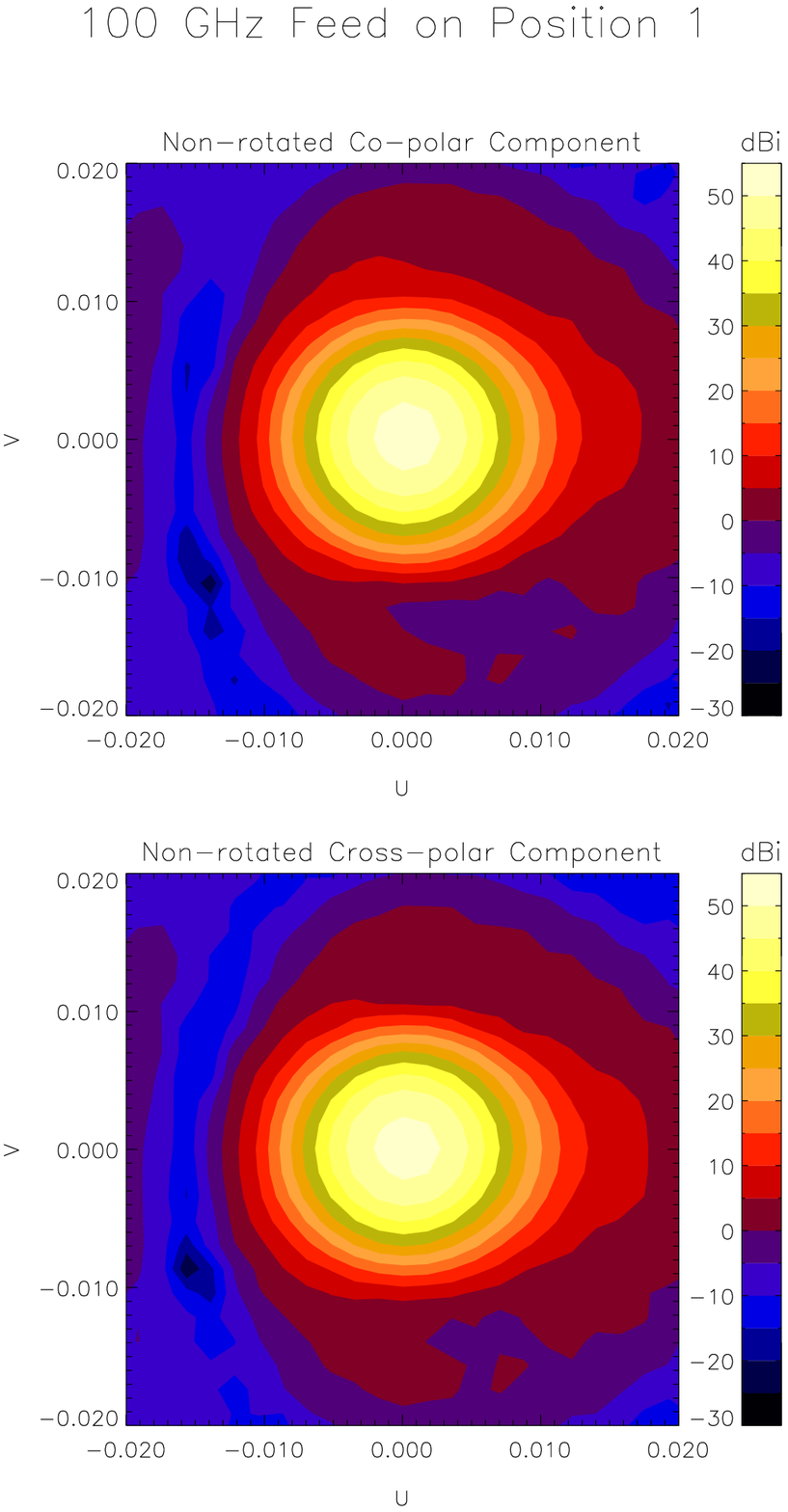,height=4.2in,clip=,silent=}}
\end{minipage}\hfill
\begin{minipage}[t]{8.5cm}
 \centerline{\psfig{file=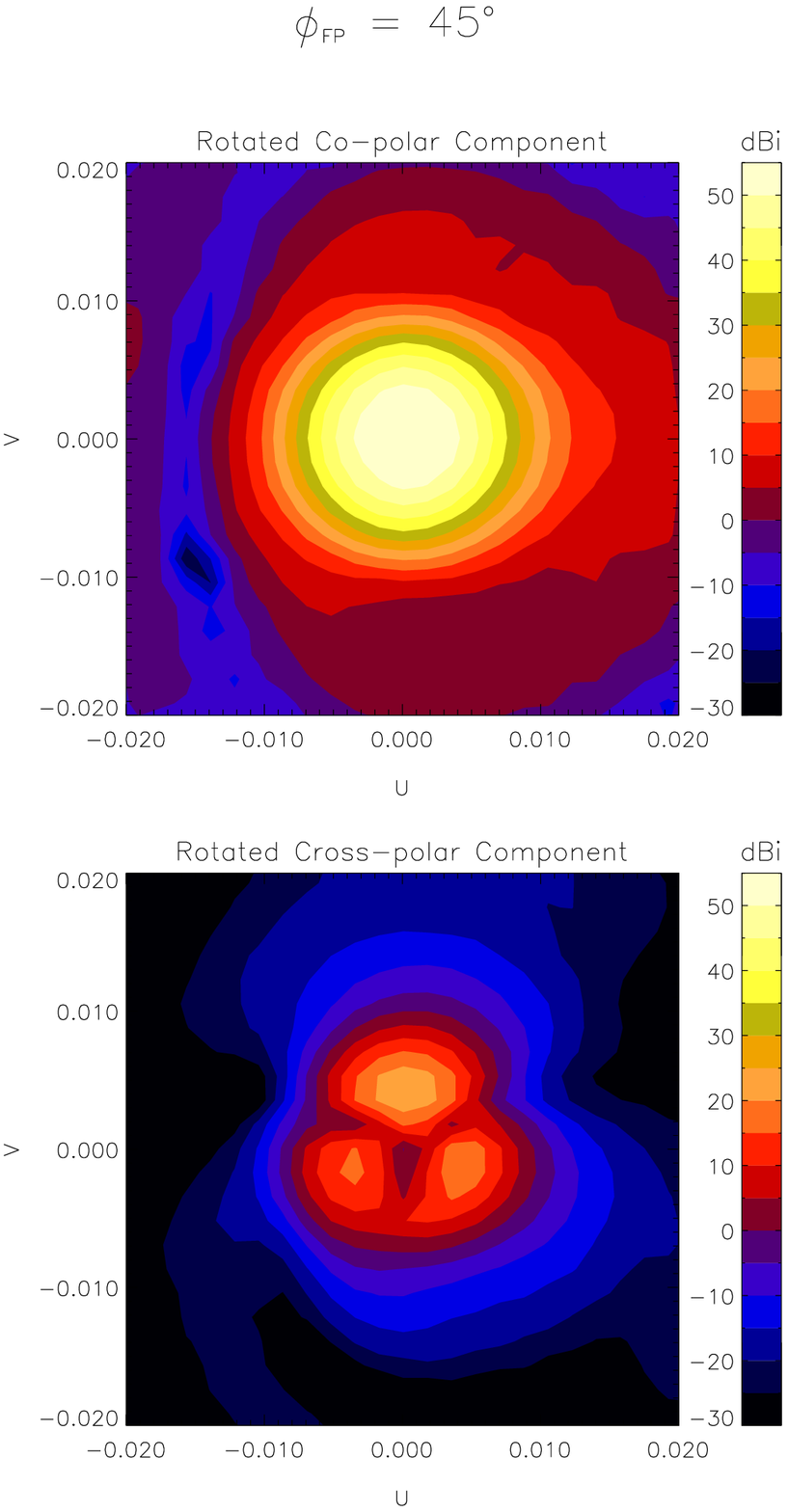,height=4.2in,clip=,silent=}}
\end{minipage}

\begin{minipage}[t]{8.5cm}
  \parbox[t]{0cm}{} \put(1.0,1.0){\small (c)} \put(10.5,1.0){\small
    (d)}
 \centerline{\psfig{file=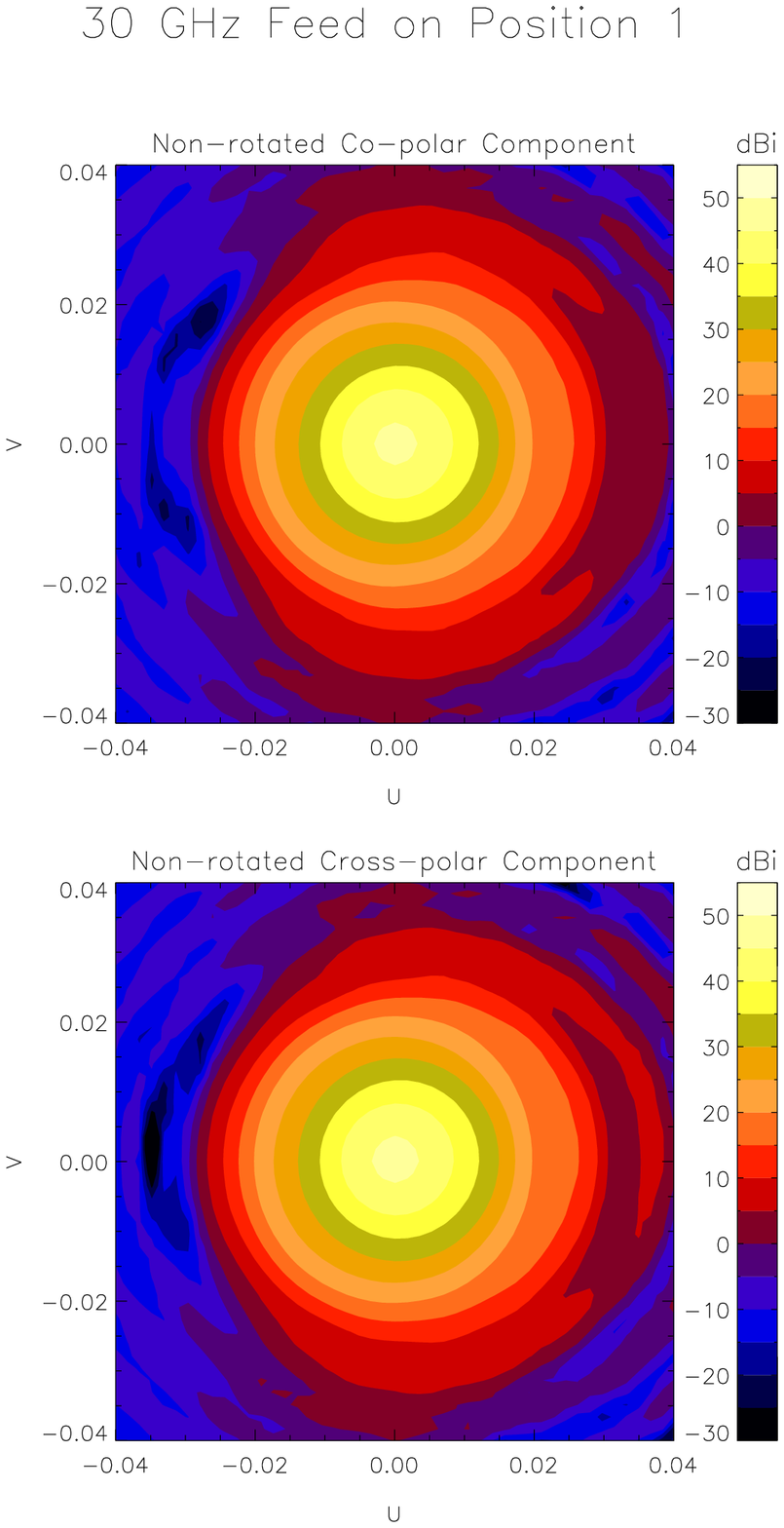,height=4.2in,clip=,silent=}}
\end{minipage}\hfill
\begin{minipage}[t]{8.5cm}
 \centerline{\psfig{file=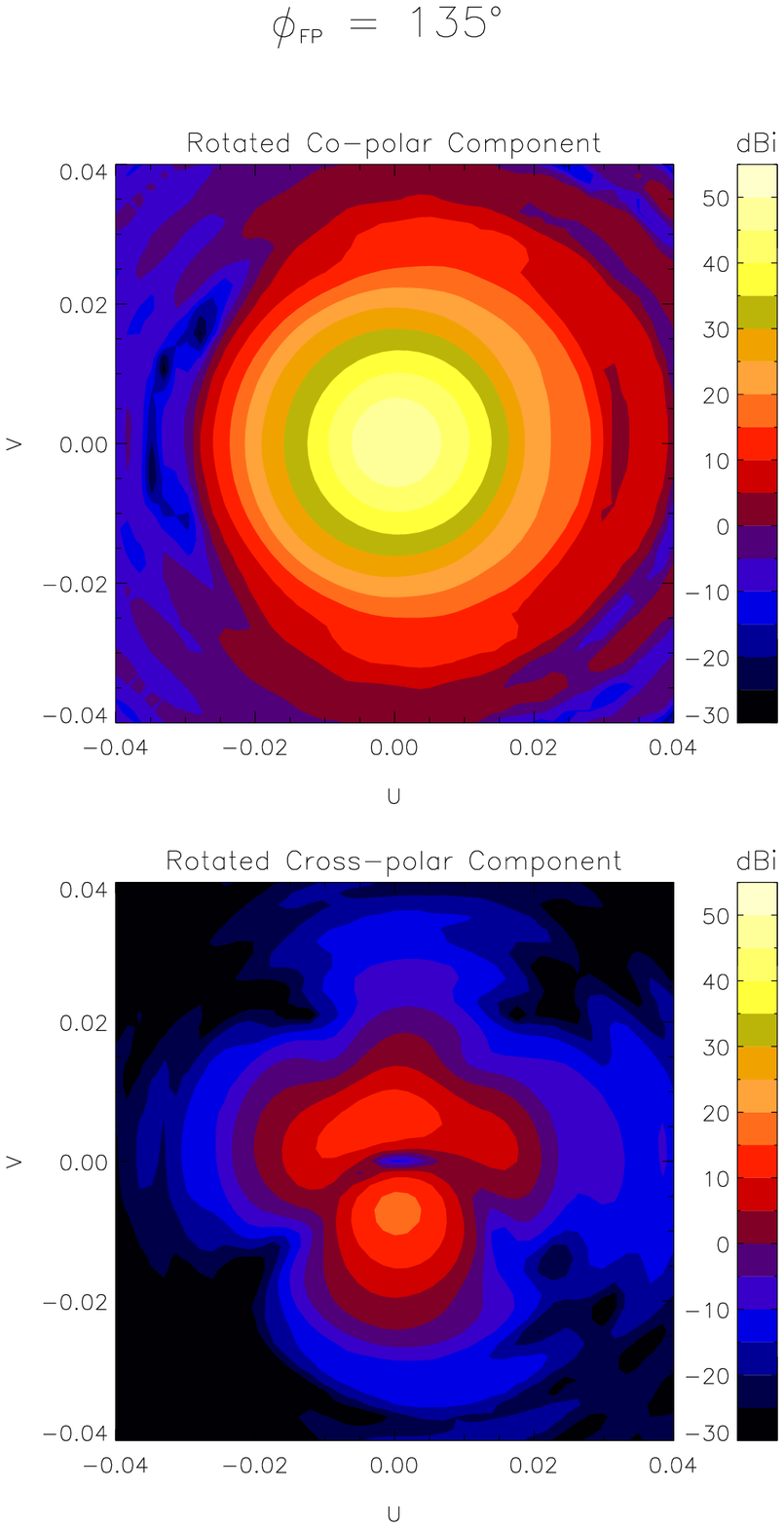,height=4.2in,clip=,silent=}}
\end{minipage}

\caption[]{Same as Figures~\ref{fig:3027psi90} and \ref{fig:1004psi0}
    for the 100~GHz feed rotated by $\psi_{FP}=45^{\circ}$ (above)
    and the 30~GHz feed rotated by $\psi_{FP}=135^{\circ}$ (below),
    both at position 1. In this figure the far field reference
    frame is rotated by (b) $\phi_{XPD}=-44.5^{\circ}$ and
    (d) $\phi_{XPD}=-134.5^{\circ}$ (see~table~\ref{tab:resultsint}).}
\label{fig:1psi45/135}
\end{figure}

As expected, for simulations with a non-rotated
feed, i.e., $\psi_{FP}=0^{\circ}$ (Figure~\ref{fig:1004psi0}), the
highest peak is at the co-polar direction before
and after the far field reference frame is rotated by $\phi_{XPD}^{nr}$
(being $\phi_{XPD}^{nr}$ always a small angle),
while for simulations with a feed orientation of $\psi_{FP}=90^{\circ}$
(Figure~\ref{fig:3027psi90}), the highest peak is at the
cross-polar direction before the far field reference
frame is rotated by $\phi_{XPD}$ which then aligns the highest peak with the
co-polar direction.
On the other hand and also as expected, for simulations with a feed
orientation of $\psi_{FP}=45^{\circ},135^{\circ}$
(Figure~\ref{fig:1psi45/135}), the co- and cross-polar
patterns have similar peaks before the rotation of the far field
reference frame by $\phi_{XPD}$.

\end{document}